\documentclass[superscriptaddress,amsmath,amssymb,aps,prd,preprintnumbers,
               showpacs,twocolumn,nofootinbib]{revtex4-2}

\usepackage[T1]{fontenc}
\usepackage[utf8]{inputenc}
\usepackage{amsmath,amssymb,bm,mathtools}
\usepackage{booktabs}
\usepackage{balance}
\usepackage[colorlinks=true,pdfstartview=Fit,linkcolor=blue,citecolor=blue,
            urlcolor=blue,breaklinks=true,hyperfootnotes=false]{hyperref}
\usepackage{orcidlink}

\usepackage{array}
\usepackage{mathrsfs}
\usepackage{amsthm,graphicx}
\graphicspath{{figures/}}

\begin{document}

\title{An unchanged null ray with anisotropic tidal dynamics \\ in Lorentz-violating Schwarzschild optics}

\author{Edilberto O. Silva\orcidlink{0000-0002-0297-5747}}
\email{edilberto.silva@ufma.br}
\affiliation{Programa de P\'os-Gradua\c c\~ao em F\'{\i}sica \& Coordena\c c\~ao do Curso de F\'{\i}sica -- Bacharelado, Universidade Federal do Maranh\~{a}o, 65085-580 S\~{a}o Lu\'{\i}s, Maranh\~{a}o, Brazil}

\author{J. L. de Oliveira\orcidlink{0009-0002-2892-2296}}
\email{jairo96.lopes@gmail.com}
\affiliation{Colegiado do Curso de Física, Universidade Estadual do Sudoeste da Bahia, Campus Juvino Oliveira, Itapetinga (BA), 45700-00, Brazil}

\author{Javier E. Gonzalez\orcidlink{0000-0002-5165-9783}}
\email{javiergonzalezs@academico.ufs.br}
\affiliation{Departamento de Física, Universidade Federal de Sergipe, São Cristovão, Sergipe, 49100-000, Brazil}

\author{João A. A. S. Reis\,\orcidlink{0000-0002-2831-5317}}
\email{joao.reis@uesb.edu.br}
\affiliation{Departamento de Ciências Exatas e Naturais, Universidade Estadual do Sudoeste da Bahia, Campus Juvino Oliveira, Itapetinga (BA), 45700-00, Brazil}

\author{L. Lisboa-Santos\orcidlink{0000-0003-4939-3856}}
\email{leticia.santos@uesb.edu.br}
\affiliation{
Departamento de Ci\^encias Exatas e Tecnológicas, Universidade Estadual do Sudoeste da Bahia, Campus Vitória da Conquista, Vitória da Conquista (BA), 45031-900, Brazil
}
\affiliation{Programa de P\'os-Gradua\c{c}\~{a}o em F\'{\i}sica, Universidade Federal do Maranh\~{a}o, Campus Universit\'{a}rio do Bacanga, S\~{a}o Lu\'{i}s (MA), 65085-580, Brazil}

\date{\today}

\begin{abstract}
We investigate the transverse dynamics of Lorentz-violating
photon congruences in a prescribed Schwarzschild background.
Within a rank-one nonbirefringent sector of the Standard-Model
Extension, we identify a radial photon ray that remains an
exact affine null geodesic of both the gravitational and optical
metrics, despite their distinct transverse tidal responses.
We derive the exact optical tidal matrix and a material-frame
Jacobi map, obtaining a Volterra representation, a closed
first-order asymptotic deformation coefficient, and a uniform
error bound. Generalizing the preferred-field configuration
to arbitrary radial profiles, we show that identical on-ray
propagation can coexist with different transverse deformations
controlled by off-ray gradients. These results establish a
geometrical distinction between the preservation of individual
photon trajectories and the evolution of neighboring rays
within Lorentz-violating optical geometries.
\end{abstract}

\maketitle

\section{Introduction}
\label{sec:introduction}

The Standard-Model Extension (SME) provides an effective-field-theory
framework for investigating possible violations of local Lorentz and
CPT symmetries in particle physics and gravity
\cite{KosteleckySamuel1989,ColladayKostelecky1997,
ColladayKostelecky1998,Kostelecky2004Gravity}.
Experimental searches and their theoretical interpretation have been
extensively reviewed \cite{Mattingly2005,JacobsonLiberatiMattingly2006,
Liberati2013,KosteleckyRussell2011}.
The photon sector is particularly sensitive to these modifications:
Lorentz-violating coefficients can alter the dispersion relation,
polarization properties, and characteristic surfaces of electromagnetic
waves \cite{KosteleckyMewes2001,KosteleckyMewes2002,KosteleckyMewes2009}.
In a curved background, the distinction between photon characteristics
and the geometry measured by conventional matter acquires a direct
kinematical significance.

Effective optical geometries have a long history, from Gordon's
metric description of light in moving media and the geometrical
interpretation of electromagnetic fields in gravitation to modern
constitutive approaches
\cite{Gordon1923,Plebanski1960,HehlObukhov2003,Itin2009}.
Related effective-metric constructions arise in nonlinear
electrodynamics and transformation optics
\cite{NovelloDeLorenciSalimKlippert2000,LeonhardtPhilbin2006}.
For Lorentz-violating photons, Betschart, Kant, and Klinkhamer studied
nonbirefringent modified Maxwell theory in Schwarzschild spacetime,
including rank-one effective metrics and nonvanishing optical Ricci
curvature on a Ricci-flat gravitational background
\cite{BetschartKantKlinkhamer2009}.
Subsequent analyses addressed eikonal photon dynamics, optical metrics
in general linear media, and the factorized characteristics of
dyadic Lorentz-violating electrodynamics
\cite{Schreck2015Eikonal,Thompson2018OpticalMetric,
GoulartFelipeOtoni2026}.
The use of an optical metric is therefore an established starting
point, not a novelty claimed by the present work.

The relation between distinct metric structures has also been
studied through conformal and disformal changes of geometry
\cite{Bekenstein1993,BettoniLiberati2013,BenAchour2025DHOST}.
Those constructions differ in their dynamical ingredients, and
results obtained for scalar--tensor systems cannot be identified
without qualification with a prescribed spacelike vector in the
SME photon sector. They nevertheless help distinguish properties
of individual null curves from properties of the surrounding
geodesic flow.

A single trajectory generally contains less geometrical information
than a congruence of neighboring rays. The connection governs the
geodesic equation, whereas its transverse derivatives determine
the screen-projected curvature entering geodesic deviation.
The Jacobi map and optical-scalar equations provide the corresponding
description of focusing, shear, and transverse deformation
\cite{Sachs1961,SeitzSchneiderEhlers1994,Perlick2004,
ReimbergAbramo2013}.
Their use in gravitational lensing is standard
\cite{SchneiderEhlersFalco1992,LiGuoLiuZhuZhao2025};
for electrodynamics in media, a related covariant treatment was
developed by Thompson and Fathi \cite{ThompsonFathi2017}.
These results establish the kinematical tools but do not determine
the response of the Lorentz-violating optical geometry considered here.

Our central question is whether an anisotropic photon-sector
deformation can preserve a particular parametrized null ray while
changing the tidal response of the rays surrounding it.
We work with a CPT-even rank-one coefficient on a prescribed
Schwarzschild background. The associated modified Maxwell action
admits an exact nonbirefringent optical-metric representation;
classical SME kinematics and its links with Finsler-type structures
provide additional context for the worldline formulation
\cite{KosteleckyRussell2010,Kostelecky2011Finsler,
KosteleckyRussellTso2012,ReisSchreckThibes2026}.
Our contribution is not another derivation of the general optical
metric, but the explicit comparison of a shared central ray with
its nontrivial transverse Jacobi dynamics.

For a particular normalized spacelike preferred field, we identify
an outgoing equatorial radial ray that is null and affinely geodesic
for both the Schwarzschild and optical metrics. The ray retains its
Schwarzschild-coordinate travel-time relation and the gravitational
frequency ratio measured by static matter observers.
The optical tidal matrix in a parallel screen is nevertheless
\begin{equation}
\mathcal T_{AB}=\operatorname{diag}\bigl(\alpha F(r),0\bigr),
\label{eq:intro_tidal_result}
\end{equation}
where $\alpha$ measures the rank-one deformation and $F(r)>0$ is
an explicit radial function. The unequal eigenvalues produce a
polar Jacobi response without modifying the fiducial geodesic.
This provides a concrete example of the distinction between
on-ray propagation data and transverse curvature information.

To define a material-frame response, we distinguish separations
in an optical screen from lengths measured by observers coupled
to the Schwarzschild metric. Infinitesimal emission angles are
specified by local group-velocity directions, and the resulting
source-to-screen Jacobi map gives a directional beam-deformation
ratio. The polar Jacobi equation admits an exact Volterra integral
representation, a closed asymptotic coefficient, and a uniform
bound on the perturbative error. This geometrical map should not be
identified by itself with a photometric magnification or with an
image observable at one reception event; such claims require
additional source--observer information and amplitude transport.
Bilocal treatments of light propagation make this distinction
particularly clear \cite{KorzynskiUzun2024}.

We further investigate the dependence on the preferred field away
from the central curve. A family of normalized fields has identical
on-ray values and preserves the same null geodesic, but yields
different transverse gradients and hence different tidal matrices.
A general positive radial profile leads to a closed expression for
the polar tidal eigenvalue and an associated inverse problem.
The construction shows that the photon coupling alone does not fix
the transverse response when the preferred background is freely
prescribed. It also separates the sign of a compressive tidal term
from the integrated conditions for conjugate-point formation.

The analysis is kinematic: the Schwarzschild metric and preferred
field are prescribed independently, and no coupled gravitational
field equations are solved. This assumption must be distinguished
from dynamical gravitational SME treatments and matter--gravity
couplings \cite{BaileyKostelecky2006,KosteleckyTasson2011}.
We work in geometrical optics and within linear geodesic deviation;
we do not compute electromagnetic fluxes or detector-specific signals.
The parameter interval supporting a Lorentzian optical metric also
does not, by itself, establish causal consistency relative to
conventional matter or evade experimental constraints.

The paper is organized as follows.
\hyperref[sec:model]{Section~\ref*{sec:model}} establishes the photon
sector, exact optical metric, and measurement conventions.
\hyperref[sec:central_ray]{Section~\ref*{sec:central_ray}} identifies
the shared null ray, while
\hyperref[sec:tidal]{Sec.~\ref*{sec:tidal}} derives its transverse
optical curvature.
\hyperref[sec:beam]{Section~\ref*{sec:beam}} develops the material
Jacobi map and asymptotic deformation, and
\hyperref[sec:profiles]{Sec.~\ref*{sec:profiles}} examines the
dependence on preferred-field gradients.
\hyperref[sec:conclusions]{Section~\ref*{sec:conclusions}} summarizes
the results and limitations.
The worldline, connection, curvature, and Volterra calculations are
detailed in
\hyperref[app:type2]{Appendices~\ref*{app:type2}}--
\hyperref[app:jacobi_asymptotic]{\ref*{app:jacobi_asymptotic}}.

\section{Optical model and physical conventions}
\label{sec:model}

We work with signature $(+,-,-,-)$ and geometrized units. The gravitational
metric $g_{\mu\nu}$ defines the local clocks and spatial measurements of
conventional matter, whereas the photon characteristics are governed by an
effective optical metric. This matter-sector convention must be specified:
without it, a comparison between photon and matter null cones is not an
operational statement.

\subsection{Rank-one photon sector and its exact optical metric}
\label{subsec:model_rank_one}

We consider a prescribed unit spacelike field $n^\mu(x)$ and a constant,
dimensionless photon-sector parameter $\kappa_A$, with
\begin{equation}
 g_{\mu\nu}n^\mu n^\nu=-1.
 \label{eq:model_unit}
\end{equation}
The traceless, symmetric coefficient is
\begin{equation}
 \widetilde\kappa^{\mu\nu}
 =\kappa_A\left(n^\mu n^\nu+\frac14g^{\mu\nu}\right),
 \qquad \nabla_\rho\kappa_A=0.
 \label{eq:model_tildek}
\end{equation}
No gravitational or preferred-field equations are imposed on this profile.
The broader constitutive description of characteristic surfaces in
linear and nonlinear electrodynamics is discussed in
Refs.~\cite{ObukhovFukuiRubilar2000,ObukhovRubilar2002}.
For the CPT-even photon action of the minimal Standard-Model Extension,
we choose
\begin{align}
 S_\gamma=-\frac14\int d^4x\,\sqrt{-g}\,
 \left(g^{\mu\rho}g^{\nu\sigma}
 +(k_F)^{\mu\nu\rho\sigma}\right)
 F_{\mu\nu}F_{\rho\sigma},
 \label{eq:model_em_action}
\end{align}
where $F_{\mu\nu}=2\partial_{[\mu}A_{\nu]}$ and
\begin{align}
 (k_F)^{\mu\nu\rho\sigma}
 =\frac12\big(&g^{\mu\rho}\widetilde\kappa^{\nu\sigma}
 -g^{\mu\sigma}\widetilde\kappa^{\nu\rho}
 \nonumber\\
 &-g^{\nu\rho}\widetilde\kappa^{\mu\sigma}
 +g^{\nu\sigma}\widetilde\kappa^{\mu\rho}\big).
 \label{eq:model_kF}
\end{align}
Using the antisymmetry of $F_{\mu\nu}$, the action becomes, without
expanding in $\kappa_A$,
\begin{align}
 S_\gamma=-\frac14\int d^4x\,\sqrt{-g}\,
 \big[&a\,g^{\mu\rho}g^{\nu\sigma}
 \nonumber\\[-2pt]
 &+2\kappa_A n^\mu n^\rho g^{\nu\sigma}\big]
 F_{\mu\nu}F_{\rho\sigma},
 \label{eq:model_reduced}
\end{align}
with
\begin{equation}
 a=1+\frac{\kappa_A}{2},\qquad
 b=1-\frac{\kappa_A}{2},\qquad
 -2<\kappa_A<2.
 \label{eq:model_ab}
\end{equation}
The interval ensures $a>0$ and $b>0$; it is the regular Lorentzian
domain of the optical metric used below, not a substitute for empirical
bounds on Lorentz violation \cite{KosteleckyRussell2011}.
It is also not a causality criterion relative to conventional matter.
Indeed, the local phase and group speed parallel to the preferred
axis equals $\sqrt{b/a}$, which is subluminal for $\kappa_A>0$
and exceeds the limiting matter speed when $\kappa_A<0$.
The negative-coupling results below are therefore mathematical
statements about the prescribed hyperbolic optical model, not an
independent demonstration of causal consistency for its coupling
to an otherwise Lorentz-invariant matter sector
\cite{BetschartKantKlinkhamer2009}.

A convenient constant-conformal representative is
\begin{equation}
 \widehat\Omega_{\mu\nu}
 =g_{\mu\nu}-\beta n_\mu n_\nu,
 \qquad \beta=\frac{\kappa_A}{b}.
 \label{eq:model_hat}
\end{equation}
Its exact inverse and volume-element ratio are
\begin{align}
 \widehat\Omega^{\mu\nu}
 &=g^{\mu\nu}+\alpha n^\mu n^\nu,
 &\alpha&=\frac{\kappa_A}{a}=\frac{\beta}{1+\beta},
 \label{eq:model_hat_inverse}\\
 \sqrt{-\det\widehat\Omega}
 &=\sqrt{\frac{a}{b}}\sqrt{-g},
 &1+\beta&=\frac{a}{b}>0.
 \label{eq:model_volume}
\end{align}
Here $g$ denotes $\det(g_{\mu\nu})$. The antisymmetry of the field
strength implies the exact contraction identity
\begin{align}
 &\widehat\Omega^{\mu\rho}
 \widehat\Omega^{\nu\sigma}F_{\mu\nu}F_{\rho\sigma}
 \nonumber\\
 &\quad =g^{\mu\rho}g^{\nu\sigma}F_{\mu\nu}F_{\rho\sigma}
 +2\alpha n^\mu n^\rho g^{\nu\sigma}
 F_{\mu\nu}F_{\rho\sigma}.
 \label{eq:model_contraction_identity}
\end{align}
Combining these identities with $a\alpha=\kappa_A$ gives
\begin{align}
 S_\gamma=-\frac{\sqrt{ab}}4\int d^4x\,
 \sqrt{-\det\widehat\Omega}\,
 \widehat\Omega^{\mu\rho}
 \widehat\Omega^{\nu\sigma}F_{\mu\nu}F_{\rho\sigma}.
 \label{eq:model_maxwell_equivalence}
\end{align}
The prefactor $\sqrt{ab}$ is constant. Hence the sourceless
field equations are exactly Maxwell equations in
$\widehat\Omega_{\mu\nu}$ for the specified rank-one sector.
The two physical polarizations share its characteristic cone.
This equivalence is a known property of suitable nonbirefringent
modified Maxwell theories
\cite{BetschartKantKlinkhamer2009,Schreck2015Eikonal,
GoulartFelipeOtoni2026}; it is
not proposed as a new optical-metric construction.

\subsection{Characteristics and the type--2 worldline}
\label{subsec:model_rays}

In geometrical optics, the wave covector $q_\mu=\partial_\mu S$
obeys the eikonal equation and the corresponding quadratic
characteristic Hamiltonian:
\begin{equation}
 \widehat\Omega^{\mu\nu}q_\mu q_\nu=0,
 \qquad
 \mathcal H(x,q)=\frac12\widehat\Omega^{\mu\nu}q_\mu q_\nu.
 \label{eq:model_dispersion}
\end{equation}
We use the affine characteristic orientation
\begin{equation}
 k^\mu=\frac{dx^\mu}{ds}
 =\frac{\partial\mathcal H}{\partial q_\mu}
 =\widehat\Omega^{\mu\nu}q_\nu,
 \qquad q_\mu=\widehat\Omega_{\mu\nu}k^\nu.
 \label{eq:model_characteristic}
\end{equation}
Thus $k^\mu$ is optical-null and satisfies
$k^\nu\widehat\nabla_\nu k^\mu=0$.
The parameter $s$ is fixed up to affine transformations and will be
normalized explicitly for the Schwarzschild ray in
Sec.~\ref{sec:central_ray}.

For clarity, this covector is not the type--2 momentum. The quadratic
type--2 worldline action
\cite{ReisSchreckThibes2026} reads
\begin{equation}
 S_{\rm T2}=-\frac12\int d\lambda\,
 \mathfrak e^{-1}\widehat\Omega_{\mu\nu}
 \dot x^\mu\dot x^\nu,
 \label{eq:model_type2_action}
\end{equation}
where $\mathfrak e$ is a nonvanishing einbein. Variation with respect
to $\mathfrak e$ imposes the optical-null constraint. The type--2
momentum and its relation to the eikonal covector are
\begin{equation}
 p_\mu^{\rm T2}
 =-\frac{\partial\widetilde L_{\rm T2}}{\partial\dot x^\mu}
 =\frac1{\mathfrak e}\widehat\Omega_{\mu\nu}\dot x^\nu,
 \qquad q_\mu=-p_\mu^{\rm T2}.
 \label{eq:model_type2_sign}
\end{equation}
The full Legendre transformation, constraint algebra, and parameter
orientation are demonstrated in Appendix~\ref{app:type2}.
In particular, the parameter choice $ds/d\lambda=-\mathfrak e$
identifies the Hamilton--Jacobi characteristic tangent with the
corresponding unparametrized type--2 null curve. Keeping the two
momentum conventions distinct avoids a spurious sign in the
Hamilton equations.

\subsection{Physical angles and material measurements}
\label{subsec:model_observers}

The optical metric determines ray propagation, but its normalized
screen is not identified with the physical ruler of a conventional
matter observer. We assume matter minimally coupled to $g_{\mu\nu}$.
For a future-directed matter observer $u^\mu$ satisfying
$g_{\mu\nu}u^\mu u^\nu=1$, the positive eikonal frequency on the
chosen branch is
\begin{equation}
 \omega_{\rm mat}=q_\mu u^\mu>0.
 \label{eq:model_frequency}
\end{equation}
For a matter observer at rest with respect to the preferred field,
$u^\mu n_\mu=0$, we may choose a local orthonormal frame with
$n^{\hat a}=(0,0,0,1)$. Let $Q_{\hat\imath}=-q_{\hat\imath}$
denote the spatial wave-vector components in that frame. The exact local
dispersion relation is
\begin{equation}
 \omega_{\rm mat}^2=Q_{\hat1}^2+Q_{\hat2}^2
 +\frac{b}{a}Q_{\hat3}^2.
 \label{eq:model_local_dispersion}
\end{equation}
The corresponding spatial group velocity is
\begin{equation}
 \mathbf v_{\rm grp}
 =\nabla_{\mathbf Q}\omega_{\rm mat}
 =\frac{1}{\omega_{\rm mat}}
 \left(Q_{\hat1},Q_{\hat2},\frac{b}{a}Q_{\hat3}\right),
 \label{eq:model_group_velocity}
\end{equation}
whereas the wave-normal direction is parallel to $\mathbf Q$.
Unless $\kappa_A=0$ or propagation is purely parallel or
perpendicular to the preferred axis, the two directions need not
coincide. We therefore use small \emph{group-direction} angles to
specify an emitted pencil, rather than angles between wave covectors.

In Schwarzschild coordinates, a static matter observer has
$u^\mu=f^{-1/2}\delta^\mu_t$, with $f=1-2M/r$.
For infinitesimal departures from an outgoing radial equatorial ray,
its local transverse-to-radial group-direction ratios give
\begin{align}
 d\vartheta_\theta^{\rm grp}
 &=r_o\sqrt{f_o}\,d[\theta'(r_o)],
 \label{eq:model_group_theta}\\
 d\vartheta_\phi^{\rm grp}
 &=r_o\sqrt{f_o}\,d[\phi'(r_o)],
 \label{eq:model_group_phi}
\end{align}
where $f_o=f(r_o)$ and primes denote derivatives with respect to $r$
at the vertex. To the same linear order, the transverse material
separations at a static detector on the equator are
\begin{equation}
 d\ell_\theta=r\,d\theta,
 \qquad d\ell_\phi=r\,d\phi.
 \label{eq:model_material_lengths}
\end{equation}
These conventions fix the normalization of the Jacobi map used in
Sec.~\ref{sec:beam}. In particular, a polarization-independent
optical characteristic cone does not imply that group directions
and material screen lengths coincide with wave-normal directions
and optical screen lengths.

The spacetime metric and the Lorentz-violating field are prescribed;
no dynamical equation selecting $n^\mu(x)$ has been imposed
\cite{Kostelecky2004Gravity}. We restrict the calculation to the
Schwarzschild exterior, geometrical optics, and infinitesimal
geodesic deviation. A calculation of electromagnetic flux would
further require amplitude transport, source normalization, and a
detector model; none is inferred from the Jacobi map alone.

\section{An unchanged ray in an anisotropic Schwarzschild optical geometry}
\label{sec:central_ray}

We now specialize the optical model of Sec.~\ref{sec:model} to a
prescribed Schwarzschild exterior. Our aim is to identify a null ray
that is shared by the gravitational and optical metrics, while keeping
the full off-ray geometry needed for the transverse calculation.
Throughout, $M>0$, $-2<\kappa_A<2$, and the metric signature is
$(+,-,-,-)$ \cite{Wald1984,Poisson2004}.

\subsection{Background geometry and preferred direction}
\label{subsec:central_geometry}

In Schwarzschild coordinates, the gravitational line element is
\begin{eqnarray}
\label{eq:central_schwarzschild}
 ds_g^2&=&f\,dt^2-f^{-1}dr^2
 -r^2(d\theta^2+\sin^2\theta\,d\phi^2),\notag\\
 \qquad f(r)&=&1-\frac{2M}{r},\quad r>2M.
\end{eqnarray}
We prescribe the unit spacelike field
\begin{eqnarray}
\label{eq:central_n}
 n^\mu &=& \left(0,\sqrt f\cos\theta,-\frac{\sin\theta}{r},0\right),\\
 n_\mu dx^\mu &=& -\frac{\cos\theta}{\sqrt f}\,dr
 +r\sin\theta\,d\theta.
\end{eqnarray}
Indeed,
\begin{equation}
\label{eq:central_unit_check}
 g_{\mu\nu}n^\mu n^\nu
 =-\cos^2\theta-\sin^2\theta=-1.
\end{equation}
As $r\to\infty$, the corresponding vector approaches
$\partial_z=\cos\theta\,\partial_r
-(\sin\theta/r)\partial_\theta$, where $z=r\cos\theta$.
Thus the preferred direction is asymptotically Cartesian,
although it is not covariantly constant in the Schwarzschild exterior.
The field is prescribed, rather than obtained from coupled
gravity--matter equations.

Using Eq.~\eqref{eq:model_hat}, the complete optical line element is
\begin{align}
\label{eq:central_optical_metric}
 d\widehat s^2={}&f\,dt^2
 -\frac{1+\beta\cos^2\theta}{f}\,dr^2
 +\frac{2\beta r\sin\theta\cos\theta}{\sqrt f}
 \,dr\,d\theta\nonumber\\
 &-r^2(1+\beta\sin^2\theta)\,d\theta^2
 -r^2\sin^2\theta\,d\phi^2,
\end{align}
where $\beta=\kappa_A/(1-\kappa_A/2)$.
This metric is exact in $\kappa_A$ and $M/r$.
On the equator its nonzero diagonal components are
\begin{equation}
\label{eq:central_equatorial_components}
 \widehat\Omega_{tt}=f,\quad
 \widehat\Omega_{rr}=-f^{-1},\quad
 \widehat\Omega_{\theta\theta}=-(1+\beta)r^2,\quad
 \widehat\Omega_{\phi\phi}=-r^2.
\end{equation}
Although $\widehat\Omega_{r\theta}=0$ at $\theta=\pi/2$,
the full metric retains transverse information. In particular,
\begin{equation}
\label{eq:central_offray_derivative}
 \left.\partial_\theta\widehat\Omega_{r\theta}
 \right|_{\theta=\pi/2}=-\frac{\beta r}{\sqrt f}.
\end{equation}
The equatorial restriction is therefore imposed only after
differentiating the metric when computing its connection and
curvature in Sec.~\ref{sec:tidal} and
Appendix~\ref{app:tidal_schwarzschild}.

\subsection{An exact shared affine null geodesic}
\label{subsec:central_shared_ray}

Consider the outgoing radial equatorial curve $\gamma$ with
\begin{equation}
\label{eq:central_k}
 \theta=\frac\pi2,\qquad \phi=\phi_o,\qquad
 k^\mu=\frac{dx^\mu}{ds}=\left(f^{-1},1,0,0\right).
\end{equation}
The choice $k^r=1$ fixes the affine normalization
$s=r+\mathrm{constant}$. The tangent is Schwarzschild-null,
\begin{equation}
\label{eq:central_gravitational_null}
 g_{\mu\nu}k^\mu k^\nu=f(f^{-1})^2-f^{-1}=0,
\end{equation}
and obeys $k^\nu\nabla_\nu k^\mu=0$.
For completeness, the relevant Schwarzschild connection components are
\begin{equation}
\label{eq:central_background_Gamma}
 \Gamma^t{}_{tr}=\frac{f'}{2f},\qquad
 \Gamma^r{}_{tt}=\frac{ff'}2,\qquad
 \Gamma^r{}_{rr}=-\frac{f'}{2f},
\end{equation}
which cancel the derivatives of $k^\mu$ in the affine geodesic
equation. Here $\nabla$ denotes the Levi--Civita connection of $g$.

Along $\gamma$, the preferred covector reduces to $r\,d\theta$;
thus
\begin{eqnarray}
\label{eq:central_orthogonality}
 n_\mu k^\mu &=&0,\\
 \widehat\Omega_{\mu\nu}k^\mu k^\nu
 &=& g_{\mu\nu}k^\mu k^\nu-\beta(n_\mu k^\mu)^2=0.
\end{eqnarray}
To establish optical affine geodesicity, define
$C^\rho{}_{\mu\nu}=\widehat\Gamma^\rho{}_{\mu\nu}
-\Gamma^\rho{}_{\mu\nu}[g]$.
The exact rank-one connection identity of
Appendix~\ref{app:connection} gives, when $n_\mu k^\mu=0$,
\begin{equation}
\label{eq:central_connection_contraction}
 C^\rho{}_{\mu\nu}k^\mu k^\nu
 =-\frac{\beta}{1+\beta}\,n^\rho
 k^\mu k^\nu\nabla_\mu n_\nu.
\end{equation}
Because orthogonality holds along the entire Schwarzschild
affine geodesic,
\begin{align}
\label{eq:central_orthogonality_transport}
 0&=k^\mu\nabla_\mu(n_\nu k^\nu)\nonumber\\
  &=k^\mu k^\nu\nabla_\mu n_\nu
    +n_\nu k^\mu\nabla_\mu k^\nu
   =k^\mu k^\nu\nabla_\mu n_\nu.
\end{align}
Consequently,
\begin{equation}
\label{eq:central_exact_geodesic}
 \widehat\Omega_{\mu\nu}k^\mu k^\nu=0,
 \qquad k^\nu\widehat\nabla_\nu k^\mu=0.
\end{equation}
The same parametrized curve is therefore an affine null geodesic
of both metrics, exactly for every admissible $\kappa_A$.
This result constrains the connection only through its contraction
with two central-ray tangents; it does not require the connections,
their transverse derivatives, or their curvature projections to agree.

\begin{figure}[tbp]
\centering
\includegraphics[width=0.98\linewidth]{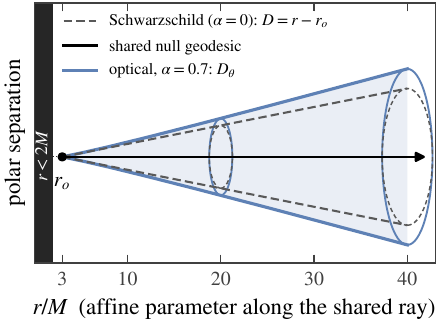}
\caption{Shared outgoing affine null geodesic (black arrow) and the
polar separation of a vertex-normalized pencil, obtained by integrating
the Jacobi equation for $p=1$, $\alpha=0.7$, and $r_o=3M$ (solid).
Dashed: the Schwarzschild ($\alpha=0$) pencil, $D=r-r_o$. The
ellipses sketch the cross sections at $r=20M$ and $40M$; the
transverse scale is magnified. The dark band marks $r<2M$; emission
and reception are restricted to $r>r_o>2M$.}
\label{fig:shared_ray_schematic}
\end{figure}

Figure~\ref{fig:shared_ray_schematic} illustrates this distinction
at the level of an infinitesimal bundle. The black central trajectory
is common to both metrics, while the surrounding optical Jacobi fields
evolve differently. The polar separations shown in the figure are obtained by
integrating the Jacobi equation derived in Sec.~\ref{sec:tidal};
for $\alpha=0$ the tidal matrix vanishes on this ray and the
Schwarzschild pencil grows exactly as $r-r_o$.

\subsection{Propagation time, frequency, and eikonal domain}
\label{subsec:central_observables}

Equation~\eqref{eq:central_k} gives the familiar outgoing radial
relation
\begin{equation}
\label{eq:central_dt_dr}
 \frac{dt}{dr}=\frac1{f(r)}.
\end{equation}
For static emission and reception radii $2M<r_o<r_d$,
the coordinate-time interval along the selected ray is
\begin{align}
\label{eq:central_travel_time}
 t_d-t_o
 &=\int_{r_o}^{r_d}\frac{du}{f(u)}\nonumber\\
 &=r_d-r_o+2M\ln\!\left(\frac{r_d-2M}{r_o-2M}\right).
\end{align}
This interval is defined in Schwarzschild time; it is not the
proper time recorded by a single local clock spanning the two events.

The optical dual of $k^\mu$ on $\gamma$ is
\begin{equation}
\label{eq:central_eikonal_covector}
 q_\mu=\widehat\Omega_{\mu\nu}k^\nu
 =\left(1,-f^{-1},0,0\right).
\end{equation}
Stationarity implies conservation of $q_t$ along the optical
characteristic. The value $q_t=1$ reflects our affine normalization;
the frequency ratio below does not depend on this choice.
A static matter observer has $u^\mu=f^{-1/2}\delta^\mu_t$ and measures
\begin{equation}
\label{eq:central_static_frequency}
 \omega_{\rm mat}=q_\mu u^\mu=\frac{q_t}{\sqrt f}.
\end{equation}
It follows that
\begin{equation}
\label{eq:central_redshift_ratio}
 \frac{\omega_d}{\omega_o}
 =\sqrt{\frac{f(r_o)}{f(r_d)}}.
\end{equation}
For this ray, the coordinate travel time and material redshift
retain their Schwarzschild expressions despite the optical anisotropy.

This agreement must not be extended to an entire wavefront by
assuming the standard radial Schwarzschild phase. Indeed, writing
$S_{\rm rad}=t-r_*(r)$ with $dr_*/dr=f^{-1}$, one obtains
\begin{equation}
\label{eq:central_eikonal_offray}
 \widehat\Omega^{\mu\nu}\partial_\mu S_{\rm rad}
 \partial_\nu S_{\rm rad}
 =\frac{\alpha\cos^2\theta}{f(r)}.
\end{equation}
For nonzero $\alpha$, this vanishes on the central equatorial ray
but not in a general neighborhood of it. The phase reproduces the
central covector without solving the modified eikonal equation
throughout the surrounding region. Neighboring optical rays must
instead be obtained from appropriate nearby characteristics,
whose transverse evolution is governed by the Jacobi system of
Sec.~\ref{sec:tidal}.

All statements in this section concern the prescribed exterior
configuration, $r>2M$. Neither the Schwarzschild-coordinate
expression for the field near $r=2M$ nor the unchanged observables
of $\gamma$ establish a regular horizon extension or identical
propagation for arbitrary directions.

\section{Exact transverse tidal dynamics}
\label{sec:tidal}

Although the Schwarzschild and optical metrics share the affinely
parametrized null ray of Sec.~\ref{sec:central_ray}, their transverse
geodesic dynamics need not agree. The distinction is quantified by the
optical tidal matrix, which involves derivatives of the connection
away from the fiducial trajectory. We construct this matrix in a
parallel screen and use it to describe infinitesimal neighboring rays.
Throughout, curvature quantities are computed with the full optical
metric before restriction to the equatorial ray. The resulting tidal
coefficients are exact in $\kappa_A$; geodesic deviation remains a
linearization in the ray separation
\cite{Sachs1961,SeitzSchneiderEhlers1994,Perlick2008}.

\subsection{Parallel optical screen and tidal curvature}
\label{subsec:tidal_screen}

For the outgoing central geodesic, $k^\mu=(f^{-1},1,0,0)$ and the
affine parameter is $r$, up to a constant shift. Choose the optical
screen basis
\begin{equation}
\label{eq:tidal_screen}
 E_\theta^\mu=\frac{\delta^\mu_\theta}{r\sqrt{1+\beta}},
 \qquad E_\phi^\mu=\frac{\delta^\mu_\phi}{r}.
\end{equation}
It satisfies
\begin{equation}
\label{eq:tidal_screen_norm}
 \widehat\Omega_{\mu\nu}E_A^\mu E_B^\nu=-\delta_{AB},
 \qquad
 \widehat\Omega_{\mu\nu}k^\mu E_A^\nu=0.
\end{equation}
This basis is also parallel propagated. Indeed, on the central ray,
\begin{equation}
\label{eq:tidal_screen_transport}
 \widehat\Gamma^\theta{}_{r\theta}
 =\widehat\Gamma^\phi{}_{r\phi}=\frac1r,
 \qquad k^\nu\widehat\nabla_\nu E_A^\mu=0.
\end{equation}
The remaining mixed coefficients that enter the transport equations,
$\widehat\Gamma^\theta{}_{t\theta}$, $\widehat\Gamma^r{}_{r\theta}$,
$\widehat\Gamma^t{}_{r\theta}$, and $\widehat\Gamma^r{}_{t\theta}$,
vanish on $\gamma$, so these two relations are sufficient.
These properties fix the screen normalization and remove spurious
rotation of its axes from the transverse evolution.

We adopt the curvature convention
$[\widehat\nabla_\mu,\widehat\nabla_\nu]V^\rho
=\widehat R^\rho{}_{\sigma\mu\nu}V^\sigma$ and define
\begin{equation}
\label{eq:tidal_def}
 \mathcal T_{AB}
 =-E_{A\rho}\widehat R^\rho{}_{\sigma\mu\nu}
   k^\sigma k^\mu E_B^\nu,
 \qquad
 E_{A\rho}=\widehat\Omega_{\rho\lambda}E_A^\lambda.
\end{equation}
The screen-index convention uses $\delta_{AB}$ to raise and lower
capital indices; the overall minus sign compensates for the negative
norm of the spacelike optical screen. With this definition, the
transverse Jacobi equation has the form $Y^{A\prime\prime}
=\mathcal T^A{}_B Y^B$.

The mixed optical component $\widehat\Omega_{r\theta}$ vanishes on
the equator, but its transverse derivative does not
[Eq.~\eqref{eq:central_offray_derivative}]. In particular, the
curvature cannot be recovered by taking derivatives of the metric
\emph{after} restricting it to the central curve. The full calculation and
an independent derivation from neighboring geodesics are given in
Appendix~\ref{app:tidal_schwarzschild}.

For the field of Eq.~\eqref{eq:central_n}, the polar projection is
\begin{equation}
\label{eq:tidal_polar_projection}
 \mathcal T_{\theta\theta}
 =\frac{1}{f^2}\widehat R^\theta{}_{tt\theta}
 +\widehat R^\theta{}_{rr\theta},
\end{equation}
and the terms independent of the deformation cancel. The resulting
exact tidal matrix is
\begin{equation}
\label{eq:tidal_exact_matrix}
 \mathcal T_{AB}
 =\begin{pmatrix}\alpha F(r)&0\\0&0\end{pmatrix},
 \qquad
 F(r)=\frac{M(1+3\sqrt f)}
 {r^3f^{3/2}(1+\sqrt f)},
\end{equation}
where $\alpha=\kappa_A/(1+\kappa_A/2)$.
The azimuthal tidal eigenvalue vanishes, whereas the polar eigenvalue
need not. Therefore, agreement of the central affine geodesics does
not imply equality of their transverse optical curvature.

\subsection{Ricci contribution, anisotropy, and limiting regimes}
\label{subsec:tidal_curvature}

The screen trace is related to the optical Ricci tensor through
\begin{equation}
\label{eq:tidal_trace_identity}
 \delta^{AB}\mathcal T_{AB}
 =-\widehat R_{\mu\nu}k^\mu k^\nu.
\end{equation}
Thus the Ricci contraction and trace-free tidal matrix are
\begin{align}
\label{eq:tidal_ricci_shear}
 \widehat R_{\mu\nu}k^\mu k^\nu
 &=-\alpha F(r),
 \\
 \mathcal T_{AB}^{\mathrm{TF}}
 &=\frac{\alpha F(r)}2
 \begin{pmatrix}1&0\\0&-1\end{pmatrix}.
\end{align}
The first expression characterizes the trace contribution to null
focusing; the second describes the anisotropic part of the tidal
source. Neither should be identified directly with the expansion or
shear of a particular pencil, which also depend on its initial data.
In particular, Ricci flatness of $g_{\mu\nu}$ does not imply
$\widehat R_{\mu\nu}k^\mu k^\nu=0$.

Since $F(r)>0$ for $M>0$ and $r>2M$, the polar eigenvalue has the
sign of $\alpha$, equivalently of $\kappa_A$ within the regular
optical domain. Positive coupling supplies a defocusing contribution
in the optical Raychaudhuri equation, while negative coupling reverses
that contribution. This sign information alone is insufficient to
establish the presence or absence of conjugate points. Moreover, the
profile analysis of Sec.~\ref{sec:profiles} shows that it is not a
universal prediction for arbitrarily prescribed preferred fields.

Expanding at fixed $M$ for $r\gg M$ gives
\begin{equation}
\label{eq:tidal_F_weak}
 F(r)=\frac{2M}{r^3}
 +\frac{11M^2}{2r^4}+\frac{13M^3}{r^5}
 +\mathcal O\!\left(\frac{M^4}{r^6}\right).
\end{equation}
In particular, the first-order weak-field and weak-coupling term is
\begin{equation}
\label{eq:tidal_weak_lorentz}
 \mathcal T_{\theta\theta}
 =\frac{2\kappa_A M}{r^3}
 +\mathcal O\!\left(
 \frac{\kappa_A M^2}{r^4},
 \frac{\kappa_A^2 M}{r^3}
 \right).
\end{equation}
The matrix vanishes when either $\kappa_A\to0$ or $M\to0$.
Although $F$ diverges as $r\to2M^+$, this exterior limit does not
establish that the prescribed preferred field admits a smooth
extension through the Schwarzschild horizon.

\subsection{Jacobi map and optical scalars}
\label{subsec:tidal_jacobi}

A transverse deviation vector can be decomposed as
$\xi_\perp^\mu=Y^A E_A^\mu$. Since the basis is parallel propagated,
the projected geodesic-deviation equation reads
\begin{equation}
\label{eq:tidal_jacobi_vector}
 \frac{d^2Y^A}{dr^2}=\mathcal T^A{}_B Y^B.
\end{equation}
The independent angular-geodesic calculation in
Appendix~\ref{app:tidal_schwarzschild} gives the same equation and
fixes its sign without relying solely on curvature conventions.
For an infinitesimal pencil emitted from a vertex at $r_o>2M$, let
$\mathsf D(r;r_o)$ be the normalized Jacobi matrix satisfying
$\mathsf D(r_o;r_o)=0$ and $\mathsf D'(r_o;r_o)=\mathsf I$.
Because the tidal matrix is diagonal in the chosen screen, so is
$\mathsf D=\operatorname{diag}(D_\theta,D_\phi)$, with
\begin{align}
\label{eq:tidal_jacobi}
 D_\theta''&=\alpha F(r)D_\theta,
 &D_\phi''&=0,
 \nonumber\\
 D_A(r_o;r_o)&=0,
 &D_A'(r_o;r_o)&=1.
\end{align}
Hence $D_\phi(r;r_o)=r-r_o$, while $D_\theta$ carries the
anisotropic optical response. Its exact integral representation is
examined in Sec.~\ref{sec:beam} and
Appendix~\ref{app:jacobi_asymptotic}.

Away from the emission vertex and before a conjugate point, the
optical deformation matrix satisfies
$\mathsf B=\mathsf D'\mathsf D^{-1}$ and the matrix Riccati equation
$\mathsf B'=\mathcal T-\mathsf B^2$. Its trace and trace-free parts
are, respectively,
\begin{align}
\label{eq:tidal_expansion}
 \Theta
 &=\frac{D_\theta'}{D_\theta}
 +\frac{D_\phi'}{D_\phi},\\
\label{eq:tidal_shear}
 \Sigma_{\theta\theta}
 &=-\Sigma_{\phi\phi}
 =\frac12\left(
 \frac{D_\theta'}{D_\theta}
 -\frac{D_\phi'}{D_\phi}\right).
\end{align}
For the point-emitted, hypersurface-orthogonal pencil, the twist
vanishes. Taking the trace of the Riccati equation gives the optical
Raychaudhuri relation
\begin{equation}
\label{eq:tidal_raychaudhuri}
 \frac{d\Theta}{dr}
 =-\frac12\Theta^2-\Sigma_{AB}\Sigma^{AB}
 +\alpha F(r),
\end{equation}
in agreement with the standard Sachs formalism
\cite{Sachs1961,Perlick2004,ThompsonFathi2017}.

To make the distinction between tidal curvature and the integrated
beam response explicit, put $\Delta r=r-r_o>0$. Near a regular
emission vertex,
\begin{align}
\label{eq:tidal_vertex_expansion}
 D_\theta(r;r_o)
 &=\Delta r+\frac{\alpha F(r_o)}6(\Delta r)^3
 +\mathcal O((\Delta r)^4),
 \\
 D_\phi(r;r_o)&=\Delta r.
\end{align}
Consequently,
\begin{align}
\label{eq:tidal_vertex_scalars}
 \Theta
 &=\frac2{\Delta r}
 +\frac{\alpha F(r_o)}3\Delta r
 +\mathcal O((\Delta r)^2),
 \\
 \Sigma_{\theta\theta}
 &=\frac{\alpha F(r_o)}6\Delta r
 +\mathcal O((\Delta r)^2).
\end{align}
Thus the anisotropic optical tidal field generates shear even
though the normalized emitted pencil starts isotropically in its
optical screen. The $2/\Delta r$ behavior is the usual vertex
divergence of the optical expansion and does not represent a
curvature singularity at $r_o$.

The Jacobi map here is normalized in an \emph{optical} screen.
Conversion to transverse lengths and group-direction launch angles
measured by conventional matter observers is a separate step,
carried out in Sec.~\ref{sec:beam}.

\section{Material-frame Jacobi map and beam deformation}
\label{sec:beam}

The tidal matrix of Sec.~\ref{sec:tidal} controls the separation of
neighboring optical rays in a parallel screen. A source-to-receiver
response additionally requires a specification of launch directions
and transverse lengths. We use the matter coupling of
Sec.~\ref{sec:model}: emission angles are defined by the local photon
group velocity, whereas transverse distances are measured with the
Schwarzschild metric. Related beam propagation in anisotropic
cosmological geometries has also been investigated
\cite{FleuryPitrouUzan2015}. The construction below is a Jacobi map
for an infinitesimal pencil, rather than a complete photometric or
single-observer imaging prescription
\cite{SeitzSchneiderEhlers1994,LiGuoLiuZhuZhao2025,KorzynskiUzun2024}.

\subsection{From the optical screen to material distances}
\label{subsec:beam_material_map}

Consider a point source static at $r_o>2M$ and a static reception
screen at $r>r_o$, centered on the outgoing equatorial ray. Let
$\eta(r)$ and $\zeta(r)$ denote first-order polar and azimuthal
coordinate displacements at fixed radial coordinate. In the
parallel-transported optical screen of Eq.~\eqref{eq:tidal_screen},
\begin{equation}
\label{eq:beam_optical_screen_displacements}
 Y_\theta=r\sqrt{1+\beta}\,\eta,
 \qquad Y_\phi=r\,\zeta.
\end{equation}
The corresponding transverse separations measured by static matter
observers are instead
\begin{equation}
\label{eq:beam_material_separations}
 d\ell_\theta=r\,\eta,
 \qquad d\ell_\phi=r\,\zeta.
\end{equation}
The distinction is essential because the optical and matter metrics
assign different lengths to the polar screen direction.

For an infinitesimal change in the initial ray direction, the
local group-velocity angles satisfy
\begin{equation}
\label{eq:beam_group_launch}
 d\vartheta_\theta^{\rm grp}
 =r_o\sqrt{f_o}\,\eta'(r_o),
 \qquad
 d\vartheta_\phi^{\rm grp}
 =r_o\sqrt{f_o}\,\zeta'(r_o),
\end{equation}
where $f_o=f(r_o)$ and a prime denotes differentiation with respect
to the central affine radial parameter. These are angles in the
matter observer's local sky, not angles between optical wave normals.
At the emission vertex $\eta(r_o)=\zeta(r_o)=0$, so
Eq.~\eqref{eq:beam_optical_screen_displacements} gives the initial
optical-screen slopes
\begin{equation}
\label{eq:beam_initial_optical_slopes}
 Y_\theta'(r_o)=\frac{\sqrt{1+\beta}}{\sqrt{f_o}}
 d\vartheta_\theta^{\rm grp},
 \qquad
 Y_\phi'(r_o)=\frac{1}{\sqrt{f_o}}
 d\vartheta_\phi^{\rm grp}.
\end{equation}

Let $D_A(r;r_o)$ be the normalized Jacobi solutions of
Eq.~\eqref{eq:tidal_jacobi}, satisfying $D_A(r_o;r_o)=0$ and
$D_A'(r_o;r_o)=1$. Linearity of the Jacobi system, followed by
conversion from optical to material lengths, gives
\begin{equation}
\label{eq:beam_material_jacobi_map}
 \begin{pmatrix}d\ell_\theta\\d\ell_\phi\end{pmatrix}
 =\frac{1}{\sqrt{f_o}}
 \begin{pmatrix}D_\theta(r;r_o)&0\\0&D_\phi(r;r_o)\end{pmatrix}
 \begin{pmatrix}
 d\vartheta_\theta^{\rm grp}\\
 d\vartheta_\phi^{\rm grp}
 \end{pmatrix}.
\end{equation}
The factor $\sqrt{1+\beta}$ cancels between the polar launch slope
and its conversion to a material length. This cancellation is a
normalization effect; the anisotropic dynamics remains in $D_\theta$.

\begin{figure}[tbp]
\centering
\includegraphics[width=0.98\linewidth]{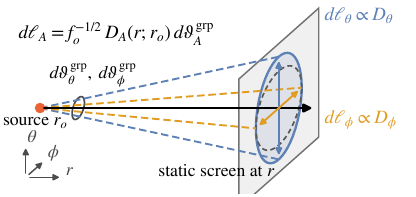}
\caption{Conceptual construction of the material-frame Jacobi map.
An infinitesimal pencil is specified by local group-direction angles
at a static source $r_o>2M$ and its transverse separation is measured
on a static reception screen at $r>r_o$. The polar and azimuthal
responses are proportional to $D_\theta$ and $D_\phi=r-r_o$,
respectively, as in Eq.~\eqref{eq:beam_material_jacobi_map}; the
departure of $D_\theta$ from $D_\phi$ is a Lorentz-violating effect,
proportional to $\alpha$ at leading order.
The drawing is not to scale and does not represent a measured flux or
an astronomical magnification.}
\label{fig:material_jacobi_schematic}
\end{figure}

The geometrical source-to-screen prescription is illustrated in
Fig.~\ref{fig:material_jacobi_schematic}. The emission angles and
reception separations are defined with the matter metric, whereas the
functions $D_A$ are obtained from optical geodesic deviation. The
polar modification of $D_\theta$ is entirely of Lorentz-violating
origin: the gravitational contributions proportional to $f'$ cancel in
the screen projection, and the deformation vanishes as $\alpha\to0$.

For an initially circular infinitesimal cone in the matter source's
angular plane, a convenient signed ratio of directional stretches is
\begin{equation}
\label{eq:beam_ratio}
 \mathcal R(r;r_o)
 \equiv\frac{D_\theta(r;r_o)}{D_\phi(r;r_o)}
 =\frac{D_\theta(r;r_o)}{r-r_o},
 \qquad r>r_o.
\end{equation}
It extends continuously to $\mathcal R(r_o;r_o)=1$. Prior to a
conjugate point, the positive ratio of major to minor semiaxes is
$\max(|\mathcal R|,|\mathcal R|^{-1})$; the signed quantity also
retains the relative orientation of the two Jacobi directions.
The material cross-sectional area per infinitesimal solid angle at
the source is
\begin{equation}
\label{eq:beam_area_map}
 d\mathcal A_{\rm mat}(r)
 =\frac{|D_\theta(r;r_o)D_\phi(r;r_o)|}{f_o}
 \,d\Omega_o^{\rm grp}.
\end{equation}
This area is defined on the prescribed reception screen. Obtaining
an observed flux, luminosity distance, or image-shape map would
require additional source--observer and amplitude-transport data
\cite{KorzynskiUzun2024}.

\subsection{Exact Volterra equation and finite-distance response}
\label{subsec:beam_volterra}

The azimuthal Jacobi equation gives
\begin{equation}
\label{eq:beam_Dphi_exact}
 D_\phi(r;r_o)=r-r_o.
\end{equation}
The polar equation $D_\theta''=\alpha F D_\theta$ can be integrated
twice using the vertex data to obtain
\begin{align}
\label{eq:beam_volterra_exact}
 D_\theta(r;r_o)
 ={}&r-r_o\nonumber\\
 &+\alpha\int_{r_o}^{r}(r-u)F(u)
 D_\theta(u;r_o)\,du.
\end{align}
This Volterra equation is exact for the linear Jacobi problem and
retains the full radial dependence of the optical tidal coefficient.
It does not assume that $|\alpha|$ or $M/r$ is small.

One iteration of Eq.~\eqref{eq:beam_volterra_exact} gives
\begin{align}
\label{eq:beam_ratio_first_order}
 \mathcal R(r;r_o)
 &=1+\alpha\mathcal I(r;r_o)+O(\alpha^2),
 \\
\label{eq:beam_I_finite}
 \mathcal I(r;r_o)
 &=\frac{1}{r-r_o}
 \int_{r_o}^{r}(r-u)(u-r_o)F(u)\,du.
\end{align}
Since $F(u)>0$ in the Schwarzschild exterior,
$\mathcal I(r;r_o)>0$ for $r>r_o$. Differentiation gives
\begin{equation}
\label{eq:beam_I_derivative}
 \frac{\partial\mathcal I}{\partial r}
 =\frac{1}{(r-r_o)^2}
 \int_{r_o}^{r}(u-r_o)^2F(u)\,du>0.
\end{equation}
The first-order accumulated response therefore increases with the
reception radius. Close to a regular vertex, $\Delta r=r-r_o>0$,
\begin{equation}
\label{eq:beam_near_vertex}
 \mathcal R(r;r_o)
 =1+\frac{\alpha F(r_o)}{6}(\Delta r)^2
 +O((\Delta r)^3).
\end{equation}
The quadratic coefficient in this local distance expansion is exact
in $\alpha$. In particular, an initially isotropic material emission
cone acquires a directional distortion continuously away from the
source.

\subsection{Asymptotic coefficient and uniform error control}
\label{subsec:beam_asymptotic}

For the preferred field of Sec.~\ref{sec:central_ray} (the case
$p=1$ of the family introduced in Sec.~\ref{sec:profiles}), the
exact tidal profile satisfies $F(u)=2M/u^3+O(M^2/u^4)$ as
$u\to\infty$. The integrated coefficient
\begin{equation}
\label{eq:beam_J_definition}
 \mathcal J(r_o)
 \equiv\int_{r_o}^{\infty}(u-r_o)F(u)\,du
\end{equation}
is therefore finite and strictly positive for every fixed $r_o>2M$,
and
\begin{equation}
\label{eq:beam_I_limit}
 \lim_{r\to\infty}\mathcal I(r;r_o)=\mathcal J(r_o).
\end{equation}
Introducing $z_o=\sqrt{1-2M/r_o}$, the integral can be evaluated
in closed form:
\begin{align}
\label{eq:beam_J_closed}
 \mathcal J(r_o)
 ={}&\frac{3-z_o}{1+z_o}
 +2\ln\left(\frac{1+z_o}{2}\right)\nonumber\\
 &+\frac{2z_o^2}{1-z_o^2}\ln z_o,
 \qquad 0<z_o<1.
\end{align}
Its weak-field expansion is
\begin{equation}
\label{eq:beam_J_weak}
 \mathcal J(r_o)
 =\frac{M}{r_o}
 +\frac{11M^2}{12r_o^2}
 +\frac{13M^3}{12r_o^3}
 +O\left(\frac{M^4}{r_o^4}\right).
\end{equation}
The closed expression determines the coefficient of the term linear
in $\alpha$, not the full finite-coupling beam deformation.

For fixed $r_o$, the exact normalized polar Jacobi function has a
finite asymptotic directional ratio. More precisely, the derivative
$D_\theta'(r)$ approaches a constant as $r\to\infty$, and
\begin{align}
\label{eq:beam_ratio_asymptotic}
 \mathcal R_\infty(r_o)
 &\equiv\lim_{r\to\infty}\mathcal R(r;r_o)\nonumber\\
 &=1+\alpha\mathcal J(r_o)+O(\alpha^2).
\end{align}
The exact asymptotic value is specified by
$1+\alpha\int_{r_o}^{\infty}F(u)D_\theta(u;r_o)\,du$;
Appendix~\ref{app:jacobi_asymptotic} proves its existence and
evaluates the first-order coefficient.

\begin{figure}[tbp]
\centering
\includegraphics[width=0.98\linewidth]{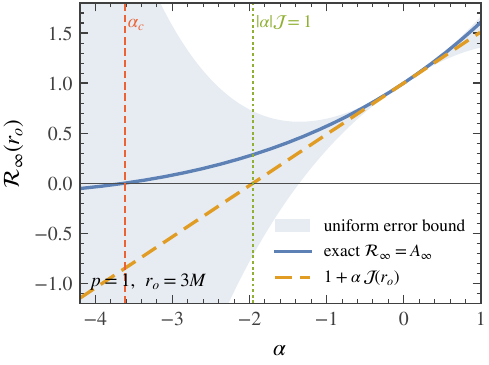}
\caption{Exact asymptotic directional ratio
$\mathcal R_\infty=A_\infty$ for $p=1$ and $r_o=3M$ (solid), obtained
numerically, compared with the first-order result
$1+\alpha\mathcal J(r_o)$ (dashed). The shaded band is the uniform
error bound of Eq.~\eqref{eq:beam_uniform_error}. The dotted line marks
$|\alpha|\mathcal J(r_o)=1$, and the dashed line $\alpha_c$ marks the
exact onset of a conjugate point.}
\label{fig:first_order_validity}
\end{figure}

The Volterra iteration also gives an error estimate uniform in the
reception radius:
\begin{align}
\label{eq:beam_uniform_error}
 \big|\mathcal R(r;r_o)-1
 -\alpha\mathcal I(r;r_o)\big|
 \leq{}&e^{|\alpha|\mathcal J(r_o)}
 -1-|\alpha|\mathcal J(r_o),
\end{align}
for all $r\geq r_o$, with the vertex value understood by
continuity. Thus the controlled first-order regime requires
$|\alpha|\mathcal J(r_o)\ll1$. Convergence of the Volterra series
and accuracy of its first-order truncation are distinct statements.
Figure~\ref{fig:first_order_validity} compares the exact asymptotic
ratio with its first-order value and with the bound
\eqref{eq:beam_uniform_error} for $r_o=3M$. For $\alpha=0.7$, for
instance, $\mathcal R_\infty=1.404$, whereas $1+\alpha\mathcal J=1.358$;
the difference $0.046$ lies within the bound $0.073$.

The coefficient is monotone in the source radius and obeys
\begin{equation}
\label{eq:beam_J_bounds}
 0<\mathcal J(r_o)<3-2\ln2,
 \qquad
 \frac{d\mathcal J}{dr_o}
 =-\int_{r_o}^{\infty}F(u)\,du<0.
\end{equation}
The finite limit $\mathcal J(2M^+)=3-2\ln2$ refers only to the
exterior integral. It does not imply a regular extension of the
preferred field, the static source, or the reception protocol
through the horizon.

\subsection{Directional behavior and conjugate-point criteria}
\label{subsec:beam_interpretation}

The exact Volterra equation provides further information without a
small-coupling expansion. When $\alpha>0$, its positive kernel
ensures $D_\theta(r;r_o)>r-r_o$ for $r>r_o$, and hence
$\mathcal R>1$. In fact,
\begin{equation}
\label{eq:beam_ratio_monotonic}
 \frac{d\mathcal R}{dr}
 =\frac{\alpha}{(r-r_o)^2}
 \int_{r_o}^{r}(u-r_o)F(u)
 D_\theta(u;r_o)\,du>0.
\end{equation}
There is no conjugate point beyond the emission vertex in this
positive-coupling case.

\begin{figure}[tbp]
\centering
\includegraphics[width=0.98\linewidth]{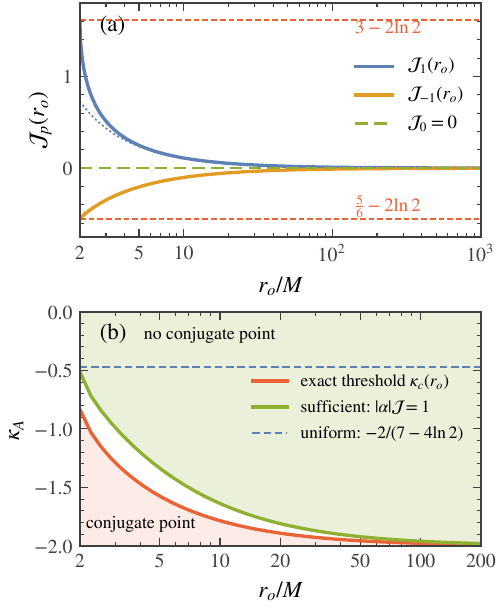}
\caption{(a) Closed-form asymptotic coefficients $\mathcal J_1(r_o)$
[Eq.~\eqref{eq:beam_J_closed}] and $\mathcal J_{-1}(r_o)$
[Eq.~\eqref{eq:profiles_J_minus_one}], together with their limits
$3-2\ln2$ and $5/6-2\ln2$ as $r_o\to2M^+$; dotted lines show the
two-term weak-field series. (b) Conjugate points for $p=1$ and
$\kappa_A<0$: exact threshold $\kappa_c(r_o)$ obtained numerically,
the sufficient condition $|\alpha|\mathcal J(r_o)=1$, and the uniform
bound $-2/(7-4\ln2)$. The green region is covered by the sufficient
condition, and the white band is free of conjugate points but not
covered by it.}
\label{fig:coefficients_conjugate}
\end{figure}

For $\alpha<0$, the polar tidal acceleration is compressive, but its
sign does not by itself prove the formation of a conjugate point
\cite{Penrose1965,Perlick2008}. As shown in
Appendix~\ref{app:jacobi_asymptotic}, a sufficient condition for
excluding a later zero of the polar Jacobi function is
\begin{equation}
\label{eq:beam_no_caustic_sufficient}
 |\alpha|\mathcal J(r_o)<1.
\end{equation}
Since $\mathcal J(r_o)<3-2\ln2$ for all $r_o>2M$, a sufficient
condition independent of the emission radius follows:
\begin{equation}
\label{eq:beam_no_caustic_uniform}
 -\frac{2}{7-4\ln2}<\kappa_A<0.
\end{equation}
These criteria are sufficient, not necessary; they do not classify
all negative couplings. A direct numerical integration of the polar
Jacobi equation locates the exact threshold $\kappa_c(r_o)$ below
which a conjugate point forms on the outgoing ray. As shown in
Fig.~\ref{fig:coefficients_conjugate}(b), the sufficient condition
\eqref{eq:beam_no_caustic_sufficient} is conservative: at $r_o=3M$
it guarantees the absence of conjugate points for $\kappa_A>-0.988$,
whereas the exact threshold is $\kappa_c\simeq-1.288$
($\alpha_c\simeq-3.62$).

 The negative-coupling domain also requires
the causal qualification relative to conventional matter stated in
Sec.~\ref{sec:model}.

The results above concern the directional stretching and material
area response of a specified infinitesimal photon pencil. A different
choice of emission angles, observer motion, or reception protocol
requires a correspondingly modified source-to-screen map. In
particular, the present Jacobi determinant is not by itself an
experimental magnification or photon flux.

\section{Dependence on preferred-field gradients}
\label{sec:profiles}

The optical tidal response obtained above is not determined by the
value of the preferred field on the fiducial ray alone. In a rank-one
optical geometry, transverse derivatives of that field enter the
connection and its curvature. We make this dependence explicit while
holding the Schwarzschild metric and the constant photon coefficient
$\kappa_A$ fixed. All backgrounds considered below are prescribed;
we do not impose a field equation selecting their radial profiles.

\subsection{A family with identical on-ray data}
\label{subsec:profiles_family}

For a fixed real parameter $p$, let
\begin{eqnarray}
 n_\mu^{(p)}dx^\mu &=& \frac{-a_p(r)\cos\theta\,dr+r\sin\theta\,d\theta}{N_p(r,\theta)},\\
 a_p(r)&=&f(r)^{-p/2},\label{eq:profiles_family}\\
 \notag\\
 N_p(r,\theta)^2 &=& f(r)a_p(r)^2\cos^2\theta+\sin^2\theta \notag\\
 &=&f(r)^{1-p}\cos^2\theta+\sin^2\theta. \label{eq:profiles_normalization}
\end{eqnarray}
Since $f>0$ for $r>2M$, the denominator is positive and the
field is smooth throughout the Schwarzschild exterior. Direct
contraction gives
\begin{equation}
 g^{\mu\nu}n_\mu^{(p)}n_\nu^{(p)}
 =-\frac{fa_p^2\cos^2\theta+\sin^2\theta}{N_p^2}=-1.
 \label{eq:profiles_unit_check}
\end{equation}
For every fixed finite $p$, the preferred direction approaches the
same Cartesian direction as $r\to\infty$. In particular, $p=1$
recovers Eq.~\eqref{eq:central_n} exactly.

On the equatorial radial ray, all profiles have the same covector:
\begin{equation}
 \left.n_\mu^{(p)}dx^\mu\right|_{\theta=\pi/2}
 =r\,d\theta,
 \qquad
 \left.n_\mu^{(p)}k^\mu\right|_{\theta=\pi/2}=0,
 \label{eq:profiles_onray}
\end{equation}
where $k^\mu=(f^{-1},1,0,0)$. The exact argument of
Sec.~\ref{sec:central_ray} consequently establishes that this
curve is an affinely parametrized null geodesic of both
$g_{\mu\nu}$ and
$\widehat\Omega_{\mu\nu}^{(p)}
=g_{\mu\nu}-\beta n_\mu^{(p)}n_\nu^{(p)}$ for every $p$.
The coordinate travel time and static-matter frequency ratio of the
central ray therefore remain the same across the family.

The off-ray data, by contrast, are different:
\begin{equation}
 \left.\partial_\theta n_r^{(p)}\right|_{\theta=\pi/2}
 =a_p(r)=f(r)^{-p/2}.
 \label{eq:profiles_gradient_onray}
\end{equation}
Thus $p$ distinguishes physically different prescribed fields, not
alternative coordinates or gauge descriptions of the same field.

\subsection{Exact tidal response and sign classification}
\label{subsec:profiles_tidal}

The corresponding optical line element is
\begin{equation}
 d\widehat s_p^2=ds_g^2
 -\frac{\beta}{N_p^2}
 \left(-a_p\cos\theta\,dr+r\sin\theta\,d\theta\right)^2.
 \label{eq:profiles_full_metric}
\end{equation}
To retain its transverse information, we expand only after forming
the full optical metric. Writing $\eta=\theta-\pi/2$, the terms
needed to linearize the polar geodesic equation are
\begin{align}
 \widehat\Omega_{rr}^{(p)}
 &=-f^{-1}-\beta a_p^2\eta^2+O(\eta^4),
 \nonumber\\
 \widehat\Omega_{r\theta}^{(p)}
 &=-\beta r a_p\eta+O(\eta^3),
 \nonumber\\
 \widehat\Omega_{\theta\theta}^{(p)}
 &=-(1+\beta)r^2+\beta r^2fa_p^2\eta^2+O(\eta^4).
 \label{eq:profiles_metric_expansion}
\end{align}
The linearized geodesic equation becomes
\begin{equation}
 \eta''+\frac{2}{r}\eta'
 -\frac{\alpha}{r^2}
 \bigl(a_p^2-a_p-ra_p'\bigr)\eta=0.
 \label{eq:profiles_angle_jacobi}
\end{equation}
A separate derivation from the quadratic transverse geodesic
action is supplied in Appendix~\ref{app:tidal_schwarzschild}.
Since all profiles induce the same optical screen on the central ray,
$Y_\theta=r\sqrt{1+\beta}\,\eta$ and $Y_\phi=r\zeta$ remain
parallel-screen components. Consequently,
\begin{equation}
 \mathcal T_{AB}^{(p)}
 =\operatorname{diag}\!\left(
 \frac{\alpha}{r^2}\bigl[a_p^2-a_p-ra_p'\bigr],0\right).
 \label{eq:profiles_tidal_exact}
\end{equation}
This expression is exact in $\alpha$ and $M/r$ within each
prescribed metric, while the Jacobi equation is linear in the
separation. Its screen trace gives
\begin{equation}
 \widehat R_{\mu\nu}^{(p)}k^\mu k^\nu
 =-\mathcal T_{\theta\theta}^{(p)}.
 \label{eq:profiles_ricci}
\end{equation}

For $a_p=f^{-p/2}$, differentiation yields the equivalent form
\begin{equation}
 \mathcal T_{\theta\theta}^{(p)}
 =\frac{\alpha a_p}{r^2}
 \left(a_p-1+\frac{pM}{rf}\right).
 \label{eq:profiles_general_sign}
\end{equation}
Because $0<f<1$, the expression in parentheses is strictly
positive when $p>0$ and strictly negative when $p<0$. It follows that
\begin{equation}
 \operatorname{sgn}\mathcal T_{\theta\theta}^{(p)}
 =\operatorname{sgn}(\alpha p),\qquad \alpha p\ne0,
 \label{eq:profiles_sign_rule}
\end{equation}
and the tidal matrix vanishes on the central ray for $p=0$.
This is a sign classification for the specified family, not a
universal focusing rule for the SME photon sector.

\subsection{General radial profiles and the inverse tidal problem}
\label{subsec:profiles_general}

The restriction $a_p=f^{-p/2}$ is not required for preservation of
the central ray. Let $a(r)>0$ be a sufficiently smooth function and
consider the unit preferred covector
\begin{equation}
 n_\mu^{(a)}dx^\mu
 =\frac{-a(r)\cos\theta\,dr+r\sin\theta\,d\theta}
 {\sqrt{f(r)a(r)^2\cos^2\theta+\sin^2\theta}}.
 \label{eq:profiles_arbitrary_field}
\end{equation}
All such fields have the same on-ray value $r\,d\theta$, so the
same equatorial radial curve remains a common affine null geodesic.
The calculation above extends directly to
\begin{equation}
 \mathcal T_{AB}^{(a)}
 =\operatorname{diag}\!\left(
 \frac{\alpha}{r^2}
 \bigl[a(r)^2-a(r)-ra'(r)\bigr],0\right).
 \label{eq:profiles_general_tidal}
\end{equation}
An independent variation of the transverse quadratic action confirms
Eq.~\eqref{eq:profiles_general_tidal}; see
Appendix~\ref{app:tidal_schwarzschild}.

For $\alpha\ne0$, the polar tidal component vanishes identically
if and only if
\begin{equation}
 ra'=a(a-1),\qquad a(r)=\frac{1}{1-Cr},
 \label{eq:profiles_zero_tidal_family}
\end{equation}
where $C$ is constant and the expression is restricted to intervals
of regularity and positivity. Requiring $a(r)>0$ on the complete
exterior and $a(r)\to1$ as $r\to\infty$ selects $a(r)\equiv1$.
The vanishing of this particular screen-projected tidal matrix
does not imply that the full optical spacetime is flat.

Conversely, assigning a polar tidal function $T(r)$ gives the
inverse first-order equation
\begin{equation}
 ra'(r)=a(r)^2-a(r)-\frac{r^2}{\alpha}T(r).
 \label{eq:profiles_inverse_tidal}
\end{equation}
For regular data, this equation admits local solutions, subject to
positivity of $a$. Global regularity and asymptotic conditions can
restrict their continuation. The profile freedom therefore permits
a broad class of transverse responses with identical central-ray
data, but does not supply a dynamical selection principle for the
preferred field. This is a geometrical classification rather than a
unique prediction from $\kappa_A$ alone.

\subsection{Accumulated deformation and conjugate points}
\label{subsec:profiles_examples}

Three members of the power-law family illustrate the classification:
\begin{align}
 p=1:\quad \mathcal T_{\theta\theta}^{(1)}
 &=\alpha F(r),\nonumber\\
 p=0:\quad \mathcal T_{\theta\theta}^{(0)}
 &=0,\nonumber\\
 p=-1:\quad \mathcal T_{\theta\theta}^{(-1)}
 &=\frac{\alpha}{r^2}
 \left(f-\sqrt f-\frac{M}{r\sqrt f}\right).
 \label{eq:profiles_examples}
\end{align}
The $p=1$ case reproduces Secs.~\ref{sec:tidal} and~\ref{sec:beam};
for $p=0$, the normalized Jacobi functions satisfy
$D_\theta=D_\phi=r-r_o$ exactly. For $p=-1$, setting
$z=\sqrt f\in(0,1)$ gives
\begin{equation}
 f-\sqrt f-\frac{M}{r\sqrt f}
 =\frac{(z-1)(2z^2+z+1)}{2z}<0.
 \label{eq:profiles_negative_sign}
\end{equation}
Thus positive $\alpha$ produces a compressive polar tidal term for
$p=-1$, in contrast with the positive $p=1$ response.

\begin{figure}[tbp]
\centering
\includegraphics[width=0.98\linewidth]{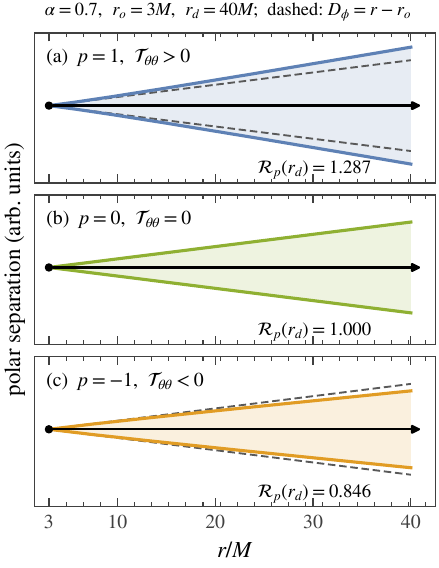}
\caption{Numerically integrated polar Jacobi separation
$D_\theta(r;r_o)$ for $a_p=f^{-p/2}$ with (a) $p=1$, (b) $p=0$, and
(c) $p=-1$, at $\alpha=0.7$ and $r_o=3M$. Dashed lines show the
azimuthal reference $D_\phi=r-r_o$. All three share the same central
ray. For $p=-1$ the pencil still expands, only more slowly than
$D_\phi$, and no conjugate point forms. The transverse scale is
magnified.}
\label{fig:preferred_profiles_schematic}
\end{figure}

Figure~\ref{fig:preferred_profiles_schematic} compares these three
cases for $\alpha>0$, using numerically integrated Jacobi solutions.
All share the same on-ray propagation, but $p=1$ yields a positive
polar tidal source, $p=0$ yields no polar Jacobi correction, and
$p=-1$ yields a negative polar source. For $p=0$ the polar width grows
exactly as $r-r_o$; for $p=-1$ the pencil still expands, only more
slowly than the azimuthal reference.

\begin{figure}[tbp]
    \centering
    \includegraphics[width=0.98\linewidth]
    {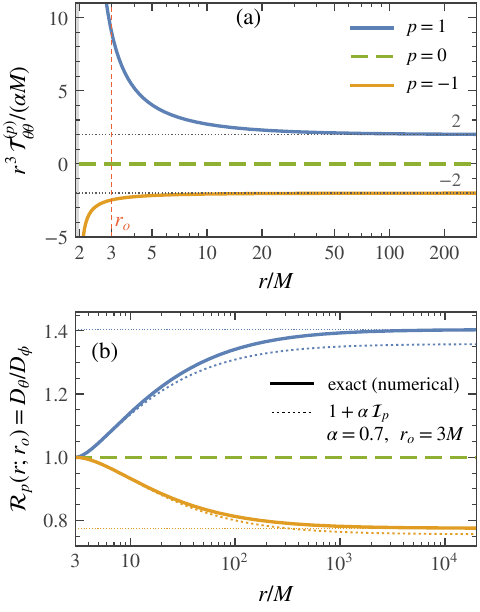}
    \caption{
    Dependence of the transverse optical dynamics on the
    preferred-field profile $a_p(r)=f(r)^{-p/2}$.
    (a) Polar tidal eigenvalue normalized to the Newtonian
    scale, $r^3\mathcal T_{\theta\theta}^{(p)}/(\alpha M)$,
    for $p=1,0,-1$; it tends to $2p$ as $r\to\infty$, and
    the dashed vertical line marks $r_o=3M$.
    (b) Directional deformation ratio
    $\mathcal R_p(r;r_o)=D_\theta(r;r_o)/D_\phi(r;r_o)$ for
    $\alpha=0.7$ and $r_o=3M$: exact numerical solution of
    the Jacobi equation with vertex initial conditions
    (solid), first-order result $1+\alpha\mathcal I_p$
    (dotted), and asymptotic values $\mathcal R_{p,\infty}$
    (thin horizontal lines).
    All three profiles preserve the same central affine
    null geodesic, despite their different transverse
    tidal responses. The plotted coupling is chosen
    to make the geometrical differences visible and
    is not intended as a phenomenological estimate.
    }
    \label{fig:profile_comparison}
\end{figure}

The influence of the preferred-field gradients on the transverse
dynamics is quantified in Fig.~\ref{fig:profile_comparison}.
For positive coupling, the three representative profiles
$p=1,0,-1$ generate positive, vanishing, and negative polar
tidal contributions, respectively. Panel (a) displays
the radial dependence of the optical tidal eigenvalue, normalized
to the Newtonian scale $\alpha M/r^3$ so that it approaches $2p$ at
large radius, while panel (b) shows the accumulated directional
deformation obtained from the Jacobi equation, together with its
first-order approximation and asymptotic values.
Although all three configurations preserve the same central
null geodesic, their neighboring rays exhibit different
transverse evolution. The strong-field enhancement near the
source and the saturation of $\mathcal R_p$ at large radius are
both visible.
In particular, a positive tidal eigenvalue produces polar
stretching, whereas a negative eigenvalue produces a
compressive contribution relative to the reference solution.
The vanishing-tidal case $p=0$ retains
$D_\theta=D_\phi=r-r_o$ exactly.
These results illustrate that the transverse response depends
not only on the Lorentz-violating coupling but also on the
off-ray gradients of the prescribed preferred field.

For fixed $p$, the weak-field expansion is
\begin{equation}
 \mathcal T_{\theta\theta}^{(p)}
 =\alpha\left[\frac{2pM}{r^3}
 +\frac{p(5p+6)M^2}{2r^4}
 +O\!\left(\frac{M^3}{r^5}\right)\right].
 \label{eq:profiles_weak}
\end{equation}
Define $F_p=\mathcal T_{\theta\theta}^{(p)}/\alpha$ by its
$\alpha$-independent expression, also at $\alpha=0$. The same
material normalization as in Sec.~\ref{sec:beam} gives the
first-order directional ratio
\begin{align}
 \mathcal R_p(r;r_o)
 &=1+\alpha\mathcal I_p(r;r_o)+O(\alpha^2),\nonumber\\
 \mathcal I_p(r;r_o)
 &=\frac{1}{r-r_o}\int_{r_o}^{r}
 (r-u)(u-r_o)F_p(u)\,du.
 \label{eq:profiles_finite_ratio}
\end{align}
Since $F_p(r)=2pM/r^3+O(r^{-4})$ for fixed $p$, the
asymptotic first-order coefficient
\begin{equation}
 \mathcal J_p(r_o)
 =\int_{r_o}^{\infty}(u-r_o)F_p(u)\,du
 \label{eq:profiles_J_definition}
\end{equation}
is finite for every $r_o>2M$. Its weak-field expansion and the
resulting asymptotic ratio are
\begin{align}
 \mathcal J_p(r_o)
 &=p\frac{M}{r_o}
 +\frac{p(5p+6)}{12}\frac{M^2}{r_o^2}
 +O\!\left(\frac{M^3}{r_o^3}\right),\nonumber\\
 \mathcal R_{p,\infty}
 &=1+\alpha\mathcal J_p(r_o)+O(\alpha^2).
 \label{eq:profiles_ratio}
\end{align}
Here $\mathcal J_1$ equals the coefficient of
Eq.~\eqref{eq:beam_J_closed}, whereas $\mathcal J_0=0$. For
$p=-1$ we also obtain the closed form
\begin{align}
 \mathcal J_{-1}(r_o)
 ={}&\frac{(1-z_o)^2(3z_o+5)}{6(1+z_o)}\nonumber\\
 &+2\ln\left(\frac{1+z_o}{2}\right),
 \qquad z_o=\sqrt{1-\frac{2M}{r_o}}.
 \label{eq:profiles_J_minus_one}
\end{align}
It is negative and bounded by [see Fig.~\ref{fig:coefficients_conjugate}(a)]
\begin{equation}
 0<-\mathcal J_{-1}(r_o)<2\ln2-\frac56<1.
 \label{eq:profiles_minus_one_bound}
\end{equation}
For $0<\kappa_A<2$, one has $0<\alpha<1$. The sufficient
no-conjugate-point criterion of
Appendix~\ref{app:jacobi_asymptotic} then applies because
$\alpha\int_{r_o}^\infty(u-r_o)|F_{-1}(u)|\,du
=-\alpha\mathcal J_{-1}(r_o)<1$.
Therefore, for $p=-1$ and positive coupling, the polar Jacobi
function remains positive and smaller than $r-r_o$ throughout the
outgoing exterior ray. A negative tidal eigenvalue consequently
does not, by itself, imply a conjugate point.

Finally, for a fixed profile whose tidal function changes sign,
the signed quantity $\mathcal J_p$ does not control the absolute
perturbative remainder. Instead, define
\begin{equation}
 \mathcal K_p(r_o)
 =\int_{r_o}^{\infty}(u-r_o)|F_p(u)|\,du<\infty.
 \label{eq:profiles_absolute_control}
\end{equation}
The Volterra series yields the uniform first-order bound
$\exp(|\alpha|\mathcal K_p)-1-|\alpha|\mathcal K_p$.
These results describe the effect of prescribed preferred-field
gradients on an infinitesimal optical congruence; they do not
establish a universal focusing law or a dynamically selected SME
background.

\section{Discussion and conclusions}
\label{sec:conclusions}

We have investigated how a Lorentz-violating photon geometry can modify
transverse propagation without changing the trajectory of a selected
light ray. The setting is a prescribed Schwarzschild exterior and a
rank-one, CPT-even photon coefficient admitting an exact effective-metric
representation. Neither effective photon metrics nor the use of Jacobi
fields in gravitational optics is new
\cite{BetschartKantKlinkhamer2009,ThompsonFathi2017,
LiGuoLiuZhuZhao2025}. The purpose of this analysis has instead been to
identify which properties of an optical congruence remain undetermined
when the motion of its central ray is known.

For the preferred field specified in Sec.~\ref{sec:central_ray}, the
outgoing equatorial radial ray is a common null geodesic with the same
affine normalization in the gravitational and optical metrics. This
statement holds exactly in the regular metric domain
$-2<\kappa_A<2$. The ray retains the Schwarzschild coordinate
travel-time relation and the standard frequency ratio measured by
static matter observers. The coincidence is restricted to this
characteristic: in particular, the conventional radial phase does not
solve the modified eikonal equation throughout a neighborhood of the
ray.

The distinction becomes apparent upon projecting the optical
curvature onto a parallel Sachs screen. The exact tidal matrix is
\begin{equation}
\label{eq:conclusions_tidal}
 \mathcal T_{AB}=\operatorname{diag}\bigl(\alpha F(r),0\bigr),
 \qquad \alpha=\frac{\kappa_A}{1+\kappa_A/2},
\end{equation}
where $F(r)>0$ is given in Eq.~\eqref{eq:tidal_exact_matrix}.
For nonzero $\alpha$, its trace and trace-free part supply,
respectively, Ricci and anisotropic contributions to the optical
deviation equation. For the normalized pencil emitted from a vertex,
the polar Jacobi function departs from the azimuthal one and a shear
develops along the outgoing ray. These are consequences of the
transverse curvature and the specified initial data; neither is
inferred from the central trajectory alone.

We have related this optical response to a material source-to-screen
map by using group-velocity launch angles and transverse lengths
measured by static observers coupled to $g_{\mu\nu}$. The signed
stretch ratio $\mathcal R=D_\theta/D_\phi$ measures the directional
deformation of that infinitesimal pencil, and its Jacobian determinant
determines the corresponding material cross-sectional area. These
quantities do not, without additional endpoint and amplitude-transport
information, define an observed flux, luminosity distance, or image
magnification \cite{KorzynskiUzun2024}. The polar Jacobi equation admits
an exact Volterra representation, and for the original preferred-field
profile its asymptotic ratio has the controlled expansion
\begin{equation}
\label{eq:conclusions_asymptotic}
 \mathcal R_\infty(r_o)
 =1+\alpha\mathcal J(r_o)+O(\alpha^2).
\end{equation}
The coefficient $\mathcal J(r_o)$ is obtained in closed form, and the
remainder is bounded uniformly in the receiver radius. The expansion
is quantitatively reliable when $|\alpha|\mathcal J(r_o)\ll1$; the
closed form of $\mathcal J$ is not an exact finite-coupling expression
for $\mathcal R_\infty$.

The preferred-field dependence provides a broader characterization of
the result. Fields that coincide along the central ray, and hence
preserve its propagation data, can have different off-ray derivatives.
For a smooth positive radial profile $a(r)$, their polar tidal
coefficient takes the exact form
\begin{equation}
\label{eq:conclusions_general_profile}
 \mathcal T_{\theta\theta}^{(a)}
 =\frac{\alpha}{r^2}\bigl[a(r)^2-a(r)-ra'(r)\bigr].
\end{equation}
In the family $a_p=f^{-p/2}$, its sign is that of $\alpha p$ for
$\alpha p\neq0$. The examples $p=1$, $p=0$, and $p=-1$ accordingly
give positive, vanishing, and negative polar tidal coefficients for
positive coupling. In the compressive $p=-1$ case, the integrated
response nevertheless excludes conjugate points beyond the vertex
throughout the positive-coupling domain. Thus the sign of a local
tidal contribution is not, by itself, a caustic criterion. More
fundamentally, the inverse profile equation of
Sec.~\ref{sec:profiles} shows that the coupling $\kappa_A$ does not
uniquely fix the tidal response when the preferred field is freely
prescribed.

The physical scope of these statements is deliberately limited. The
optical metric and its tidal coefficients are exact within the chosen
rank-one model, but the ray description assumes geometrical optics
and the Jacobi system linearizes the separation of neighboring
characteristics. The Schwarzschild geometry and preferred field are
not solutions of a coupled dynamical model established here
\cite{Kostelecky2004Gravity}; no mechanism selecting the profile has
been derived. Moreover, the Lorentzian interval
$-2<\kappa_A<2$ is not a sufficient causal admissibility condition
for conventional matter: negative coupling permits locally
superluminal group propagation along the preferred axis. Such cases
are retained as mathematical regimes of the prescribed optical model,
not asserted to be physically viable.

These qualifications distinguish the present kinematic result from a
complete phenomenological prediction. A dynamically selected
preferred background, nonradial congruences, and electromagnetic
amplitude transport would be needed to pursue that objective. Within
the stated assumptions, the calculations establish a precise example
of a more general geometric distinction: identical on-ray null motion
does not fix the transverse Jacobi response of the surrounding photon
bundle.

\begin{acknowledgments}

E. O. Silva acknowledges the support from Conselho Nacional de
Desenvolvimento Cient\'{\i}fico e Tecnol\'{o}gico (CNPq) (grant 305427/2026-1), Funda\c c\~ao de Amparo \`a Pesquisa e ao Desenvolvimento Cient\'{\i}fico e Tecnol\'{o}gico do Maranh\~ao (FAPEMA) (grant UNIVERSAL-06395/22), and Coordena\c c\~ao de Aperfei\c coamento de Pessoal de N\'{\i}vel Superior (CAPES) -- Brazil (Finance Code 001). JAASR acknowledges partial financial support from UESB through Grant AuxPPI (Edital No. 267/2024) and also gratefully recognizes support from FAPESB–CNPq/Produtividade (Grant No. 12243/2025, TOB-BOL2798/2025). J.L. de Oliveira acknowledges the Universidade Estadual do Sudoeste da Bahia (UESB) for support through its voluntary undergraduate research (IC) program.

\end{acknowledgments}

\section*{Conflict of Interest}

The authors declare that they have no competing
financial or non-financial interests that could
have influenced the research presented in this work.

\section*{Data Availability Statement}

No datasets were generated or analyzed during
the present study. This work is entirely theoretical,
and all analytical results and derivations supporting
the conclusions are presented in the manuscript.

\appendix

\section{Type--2 conventions and the quadratic optical action}
\label{app:type2}

This appendix fixes the momentum signs, constraint structure, and
parameter orientation of the massless type--2 worldline used in
Sec.~\ref{sec:model}. It also specifies the extent to which the
optical-ray description is invariant under a conformal change of
representative. The underlying electromagnetic equivalence was
established in Sec.~\ref{subsec:model_rank_one} and is not
rederived here.

\subsection{Legendre transformation and first-class constraints}
\label{app:type2_constraints}

Let $u^\mu=dx^\mu/d\lambda$ and let $\mathfrak e(\lambda)$ be a
nonvanishing einbein. For the fixed optical representative
$\widehat\Omega_{\mu\nu}$ of Eq.~\eqref{eq:model_hat}, the
quadratic type--2 action is
\begin{equation}
\label{eq:appA_action}
S_{\mathrm{T2}}=\int d\lambda\,\widetilde L,
\qquad
\widetilde L=-\frac{1}{2\mathfrak e}
\widehat\Omega_{\mu\nu}(x)u^\mu u^\nu.
\end{equation}
It is invariant under orientation-preserving reparametrizations
$\lambda\mapsto\lambda'(\lambda)$ provided that
$\mathfrak e'(\lambda')\,d\lambda'
=\mathfrak e(\lambda)\,d\lambda$.
The overall minus sign in Eq.~\eqref{eq:appA_action} is part of
the adopted type--2 convention.

The momentum defined by the type--2 prescription is the negative
of the ordinary canonical momentum of the same action:
\begin{align}
\label{eq:appA_momenta}
p_\mu^{\mathrm{T2}}
&=-\frac{\partial\widetilde L}{\partial u^\mu}
=\frac{1}{\mathfrak e}\widehat\Omega_{\mu\nu}u^\nu,
&
q_\mu
&=\frac{\partial\widetilde L}{\partial u^\mu}
=-p_\mu^{\mathrm{T2}}.
\end{align}
Inverting the velocity--momentum relation gives
\begin{equation}
\label{eq:appA_velocity}
u^\mu=\mathfrak e\widehat\Omega^{\mu\nu}
p_\nu^{\mathrm{T2}}
=-\mathfrak e\widehat\Omega^{\mu\nu}q_\nu.
\end{equation}
The type--2 Legendre transform and the ordinary canonical
Hamiltonian coincide numerically:
\begin{align}
\label{eq:appA_hamiltonian}
\widetilde H
&=-p_\mu^{\mathrm{T2}}u^\mu-\widetilde L
=-\frac{\mathfrak e}{2}
\widehat\Omega^{\mu\nu}p_\mu^{\mathrm{T2}}
p_\nu^{\mathrm{T2}},\\
H_{\mathrm c}
&=q_\mu u^\mu-\widetilde L
=-\mathfrak e\,\Phi(x,q),
\nonumber
\end{align}
where
\begin{equation}
\label{eq:appA_phi}
\Phi(x,q)=\frac12\widehat\Omega^{\mu\nu}(x)q_\mu q_\nu.
\end{equation}
The Hamiltonian is a constrained quadratic form; no
positive-definiteness is implied by the Lorentzian signature.

The canonical momentum conjugate to the einbein vanishes,
$\pi_{\mathfrak e}\approx0$. The total Hamiltonian is
$H_{\mathrm T}=-\mathfrak e\Phi+v\pi_{\mathfrak e}$, with an
arbitrary multiplier $v$. Preservation of the primary constraint
produces the secondary null constraint:
\begin{equation}
\label{eq:appA_constraint_algebra}
\dot\pi_{\mathfrak e}=\{\pi_{\mathfrak e},H_{\mathrm T}\}
=\Phi\approx0,
\qquad
\{\pi_{\mathfrak e},\Phi\}=0,
\qquad
\{\Phi,\Phi\}=0.
\end{equation}
Thus the constraints are first class, and the canonical Hamiltonian
vanishes on the constraint surface. With $(x^\mu,q_\mu)$ as
canonical variables, the nonzero fundamental brackets are
\begin{equation}
\label{eq:appA_poisson}
\{x^\mu,q_\nu\}=\delta^\mu{}_\nu,
\qquad
\{x^\mu,p_\nu^{\mathrm{T2}}\}=-\delta^\mu{}_\nu.
\end{equation}
Assigning a positive canonical bracket to $p_\mu^{\mathrm{T2}}$
while retaining Eq.~\eqref{eq:appA_momenta} would reverse the
Hamilton equations incorrectly.

The same constrained dynamics follows from the first-order action
\begin{equation}
\label{eq:appA_firstorder}
S_{\mathrm{1st}}=\int d\lambda\,
\left(q_\mu\dot x^\mu+\mathfrak e\Phi(x,q)\right),
\end{equation}
whose $q_\mu$ equation reproduces
Eq.~\eqref{eq:appA_velocity}. Eliminating $q_\mu$ returns
Eq.~\eqref{eq:appA_action} exactly.

\subsection{Null characteristics and affine orientation}
\label{app:type2_affine}

Varying the einbein in Eq.~\eqref{eq:appA_action} imposes the
optical null condition. Variation with respect to $x^\rho$
gives the geodesic equation with a nonaffine parameter:
\begin{align}
\label{eq:appA_geodesic}
\widehat\Omega_{\mu\nu}u^\mu u^\nu&=0,
&
\frac{du^\rho}{d\lambda}
+\widehat\Gamma^\rho{}_{\mu\nu}u^\mu u^\nu
&=\frac{\dot{\mathfrak e}}{\mathfrak e}u^\rho.
\end{align}
Define the affine characteristic parameter $s$ by
\begin{equation}
\label{eq:appA_parameter}
\frac{ds}{d\lambda}=-\mathfrak e.
\end{equation}
The Hamilton equations generated by $H_{\mathrm c}$ then become
\begin{align}
\label{eq:appA_characteristics}
\frac{dx^\mu}{ds}
&=\widehat\Omega^{\mu\nu}q_\nu,
&
\frac{dq_\mu}{ds}
&=-\frac12\partial_\mu\widehat\Omega^{\rho\sigma}
q_\rho q_\sigma.
\end{align}
The first equation identifies the characteristic tangent
$k^\mu=dx^\mu/ds$ with the optical dual of $q_\mu$, as in
Eq.~\eqref{eq:model_characteristic}. On the constraint surface,
these equations describe affine null geodesics of
$\widehat\Omega_{\mu\nu}$. When $q_\mu=\partial_\mu S$, they
also coincide with the Hamilton--Jacobi rays of
Eq.~\eqref{eq:model_dispersion}.

The sign in Eq.~\eqref{eq:appA_parameter} is fixed by the
chosen relation $q_\mu=-p_\mu^{\mathrm{T2}}$; it is not a
physical reversal of the photon trajectory. The sign of the
einbein and the orientation of $\lambda$ must simply be chosen
consistently with the desired future-directed characteristic
parameter $s$.

\subsection{Conformal representatives and an isotropic check}
\label{app:type2_conformal}

Let $C(x)>0$ and perform the simultaneous replacements
\begin{equation}
\label{eq:appA_conformal_action}
\widehat\Omega_{\mu\nu}\longmapsto
\widehat\Omega'_{\mu\nu}=C(x)\widehat\Omega_{\mu\nu},
\qquad
\mathfrak e\longmapsto\mathfrak e'=C(x)\mathfrak e.
\end{equation}
The type--2 Lagrangian, the canonical covector $q_\mu$, and the
first-order action are unchanged under this combined replacement.
The new optical characteristic parameter satisfies
\begin{equation}
\label{eq:appA_conformal_parameter}
\frac{ds'}{ds}=C(x(s)).
\end{equation}
Indeed, if $k^\mu=dx^\mu/ds$ is affinely parametrized for
$\widehat\Omega_{\mu\nu}$, then
$k'^\mu=dx^\mu/ds'=C^{-1}k^\mu$ is affinely parametrized
for $\widehat\Omega'_{\mu\nu}$. The unparametrized null curves
are unchanged, while a nonconstant conformal factor generally
changes the connection and the optical tidal projections.
For constant $C$, the two Levi--Civita connections and their
mixed-index Riemann tensors coincide, and affine parameters differ
at most by a constant rescaling. The calculations of the main text
use the specific representative in Eq.~\eqref{eq:model_hat}.

For comparison, an isotropic local inverse optical metric can be
written as
\begin{equation}
\label{eq:appA_isotropic_inverse}
\Omega^{\hat a\hat b}_{\mathrm{iso}}
=\operatorname{diag}\!\left(1+\widetilde\kappa_{\rm tr},
-(1-\widetilde\kappa_{\rm tr}),
-(1-\widetilde\kappa_{\rm tr}),
-(1-\widetilde\kappa_{\rm tr})\right).
\end{equation}
For $|\widetilde\kappa_{\rm tr}|<1$, a constant-conformal
covariant representative is
\begin{eqnarray}
\label{eq:appA_isotropic_line}
d\widehat s_{\mathrm{iso}}^2
&=&v_\gamma^2 f(r)dt^2-\frac{dr^2}{f(r)}\notag\\
&&-r^2(d\theta^2+\sin^2\theta\,d\phi^2),\\
v_\gamma^2&=&\frac{1-\widetilde\kappa_{\rm tr}}
{1+\widetilde\kappa_{\rm tr}}.
\end{eqnarray}
The constant coordinate transformation $T=v_\gamma t$
puts this representative into Schwarzschild form, so its
intrinsic Ricci tensor vanishes. This geometrical statement does
not identify the photon and matter clock readings: ordinary matter
continues to measure time with $g_{\mu\nu}$ as stipulated in
Sec.~\ref{sec:model}. The isotropic example is a consistency
check, not a limiting case of the anisotropic preferred-field
family of Sec.~\ref{sec:profiles}.

\section{Connection and curvature of the rank-one metric}
\label{app:connection}

This appendix collects the exact differential identities underlying the
shared-ray construction of Sec.~\ref{sec:central_ray} and the optical
curvature calculations of Sec.~\ref{sec:tidal}. In particular, the
vanishing of the difference between two geodesic accelerations on a
chosen curve must not be confused with the vanishing of the curvature
difference in directions transverse to that curve. No expansion in the
photon-sector coefficient is made.

\subsection{Inverse metric, volume element, and connection difference}

Let $n_\mu n^\mu=-1$ with indices on $n_\mu$ raised using the
gravity metric $g_{\mu\nu}$, and let $\beta$ be constant with
$1+\beta>0$. The rank-one optical metric and its inverse are
\begin{align}
\label{eq:appB_inverse}
\widehat\Omega_{\mu\nu}
 &=g_{\mu\nu}-\beta n_\mu n_\nu,
 &\widehat\Omega^{\mu\nu}
 &=g^{\mu\nu}+\alpha n^\mu n^\nu,
 &\alpha&=\frac{\beta}{1+\beta}.
\end{align}
The inverse follows by direct multiplication, while the matrix
determinant lemma gives
\begin{equation}
\label{eq:appB_determinant}
\det\widehat\Omega=(1+\beta)\det g.
\end{equation}
These relations hold for an arbitrary smooth normalized field,
regardless of whether it satisfies a dynamical equation.

Let $\nabla$ denote the Levi--Civita derivative of $g_{\mu\nu}$.
We introduce the tensors
\begin{equation}
\label{eq:appB_KSW}
 K_{\mu\nu}=\nabla_\mu n_\nu,\quad
 S_{\mu\nu}=K_{\mu\nu}+K_{\nu\mu},\quad
 W_{\mu\nu}=K_{\mu\nu}-K_{\nu\mu}.
\end{equation}
The normalization implies $n^\nu K_{\mu\nu}=0$. Since $\beta$ is
constant, metric compatibility of $\nabla$ gives
\begin{equation}
\label{eq:appB_metric_derivative}
\nabla_\mu\widehat\Omega_{\nu\sigma}
=-\beta\bigl(K_{\mu\nu}n_\sigma
+n_\nu K_{\mu\sigma}\bigr).
\end{equation}
The exact difference between the optical and gravitational
Levi--Civita connections is therefore
\begin{align}
\label{eq:appB_C_exact}
 C^\rho{}_{\mu\nu}
 &\equiv\widehat\Gamma^\rho{}_{\mu\nu}
       -\Gamma^\rho{}_{\mu\nu}[g]\nonumber\\
 &=-\frac{\beta}{2}\widehat\Omega^{\rho\sigma}
 \bigl(n_\sigma S_{\mu\nu}
 +n_\nu W_{\mu\sigma}
 +n_\mu W_{\nu\sigma}\bigr).
\end{align}
This difference is a tensor, is symmetric in its two lower
connection indices, and includes the full dependence on the
off-ray derivatives of $n_\mu$. The contraction
\begin{equation}
\label{eq:appB_n_contraction}
\widehat\Omega^{\rho\sigma}n_\sigma
=\frac{n^\rho}{1+\beta}
\end{equation}
is especially useful for evaluating its action on a tangent
orthogonal to $n^\mu$.

Because the determinant ratio in Eq.~\eqref{eq:appB_determinant}
is constant, the contracted connection difference vanishes:
\begin{equation}
\label{eq:appB_trace}
 C^\rho{}_{\rho\mu}
 =\partial_\mu\ln\left(
 \frac{\sqrt{-\det\widehat\Omega}}{\sqrt{-\det g}}
 \right)=0.
\end{equation}
Consequently, $\widehat\nabla_\mu V^\mu=\nabla_\mu V^\mu$
for any fixed vector field $V^\mu$. This identity concerns the
divergence of the \emph{same} vector field; it does not imply that
different optical and gravitational congruences have identical
expansion, shear, or Jacobi maps.

\subsection{Curvature difference and Ricci contraction}

We fix the curvature convention
\begin{equation}
\label{eq:appB_curvature_convention}
 [\nabla_\mu,\nabla_\nu]V^\rho
 =R^\rho{}_{\sigma\mu\nu}[g]V^\sigma,
 \qquad R_{\sigma\nu}[g]=R^\rho{}_{\sigma\rho\nu}[g],
\end{equation}
with identical definitions for $\widehat\nabla$.
For two torsion-free connections related by
$\widehat\nabla=\nabla+C$, the exact Riemann identity reads
\begin{align}
\label{eq:appB_R}
 \widehat R^\rho{}_{\sigma\mu\nu}
 ={}&R^\rho{}_{\sigma\mu\nu}[g]
 +\nabla_\mu C^\rho{}_{\nu\sigma}
 -\nabla_\nu C^\rho{}_{\mu\sigma}\nonumber\\
 &+C^\rho{}_{\mu\lambda}C^\lambda{}_{\nu\sigma}
 -C^\rho{}_{\nu\lambda}C^\lambda{}_{\mu\sigma}.
\end{align}
Contracting $\rho$ with the third index and using
Eq.~\eqref{eq:appB_trace} yields
\begin{equation}
\label{eq:appB_Ricci}
 \widehat R_{\mu\nu}
 =R_{\mu\nu}[g]
 +\nabla_\rho C^\rho{}_{\nu\mu}
 -C^\rho{}_{\nu\lambda}C^\lambda{}_{\rho\mu}.
\end{equation}
In particular, even if $R_{\mu\nu}[g]=0$, the optical Ricci
tensor need not vanish. The surviving terms involve both
derivatives of the connection difference and its quadratic
combinations. For any vector $k^\mu$, the relevant contraction is
\begin{align}
\label{eq:appB_Ricci_null_contraction}
 \widehat R_{\mu\nu}k^\mu k^\nu
 ={}&R_{\mu\nu}[g]k^\mu k^\nu
 +k^\mu k^\nu\nabla_\rho C^\rho{}_{\nu\mu}\nonumber\\
 &-k^\mu k^\nu C^\rho{}_{\nu\lambda}
                  C^\lambda{}_{\rho\mu}.
\end{align}
Equation~\eqref{eq:appB_Ricci_null_contraction} is one reason why
knowledge of a single geodesic is insufficient to determine the
optical tidal matrix: the curvature samples how the connection
varies in a neighborhood of the ray.

\subsection{Common affine rays and the limits of the criterion}

Let $k^\mu$ be tangent to an affinely parametrized null geodesic
of the gravitational metric, and suppose that
\begin{equation}
\label{eq:appB_common_hypotheses}
 g_{\mu\nu}k^\mu k^\nu=0,\qquad
 k^\mu\nabla_\mu k^\rho=0,\qquad
 n_\mu k^\mu=0
\end{equation}
holds along the entire curve. The rank-one form immediately implies
$\widehat\Omega_{\mu\nu}k^\mu k^\nu=0$.
Contracting Eq.~\eqref{eq:appB_C_exact} with two tangents, the
terms proportional to $n_\mu k^\mu$ vanish. Using
Eq.~\eqref{eq:appB_n_contraction}, one obtains
\begin{equation}
\label{eq:appB_contraction}
 C^\rho{}_{\mu\nu}k^\mu k^\nu
 =-\frac{\beta}{1+\beta}\,n^\rho
    k^\mu k^\nu\nabla_\mu n_\nu.
\end{equation}
Orthogonality along the curve and gravitational affine geodesicity
give, without any assumption about the derivatives of $n_\mu$
away from the curve,
\begin{align}
\label{eq:appB_orthogonality}
 0&=k^\mu\nabla_\mu(n_\nu k^\nu)\nonumber\\
  &=k^\mu k^\nu\nabla_\mu n_\nu
    +n_\nu k^\mu\nabla_\mu k^\nu
   =k^\mu k^\nu\nabla_\mu n_\nu.
\end{align}
It follows that
\begin{equation}
\label{eq:appB_shared}
 C^\rho{}_{\mu\nu}k^\mu k^\nu=0,
 \qquad k^\mu\widehat\nabla_\mu k^\rho=0.
\end{equation}
Thus the ray is optical-null and remains affinely parametrized by
the \emph{same} parameter. The conclusion is exact and requires
neither $C^\rho{}_{\mu\nu}=0$ nor equality of the curvature
tensors along the ray.

This criterion provides a useful sufficient condition for a shared
ray, not a requirement that the two metrics have the same null cone
or geodesic flow. In particular, Eq.~\eqref{eq:appB_shared}
constrains only a double contraction of $C^\rho{}_{\mu\nu}$;
its transverse derivatives and the terms entering
Eq.~\eqref{eq:appB_R} remain unrestricted by that condition.
This distinction is consistent with the broader literature on
disformal changes of geometry
\cite{Bekenstein1993,BettoniLiberati2013,BenAchour2025DHOST},
although the scalar--tensor models studied there should not be
identified with the prescribed vector background considered here.
The explicit Schwarzschild curvature projection and independent
Jacobi check are given in Appendix~\ref{app:tidal_schwarzschild}.

\section{Curvature projection and independent Jacobi derivation}
\label{app:tidal_schwarzschild}

This appendix verifies the optical tidal field by two complementary
calculations. First, we project the Riemann tensor obtained from the
\emph{full} four-dimensional optical metric. Second, we derive the
transverse equation by varying the quadratic action for neighboring
geodesics. The second method does not use the curvature tensor and
therefore provides an independent check of the tidal sign and
normalization. We retain the convention
$[\widehat\nabla_\mu,\widehat\nabla_\nu]V^\rho
=\widehat R^\rho{}_{\sigma\mu\nu}V^\sigma$
and the screen definition of Eq.~\eqref{eq:tidal_def}.

\subsection{Full-curvature projection on the common ray}
\label{app:tidal_curvature_projection}

Set $f(r)=1-2M/r$ and consider the common outgoing affine ray
$k^\mu=(f^{-1},1,0,0)$ at $\theta=\pi/2$. Its parallel optical
screen is
\begin{equation}
\label{eq:appC_screen}
 E_\theta^\mu=\frac{\delta^\mu_\theta}{r\sqrt{1+\beta}},
 \qquad
 E_\phi^\mu=\frac{\delta^\mu_\phi}{r}.
\end{equation}
Here $\widehat\Omega_{\mu\nu}E_A^\mu E_B^\nu=-\delta_{AB}$,
and the relevant screen-transport coefficients on the ray are
$\widehat\Gamma^\theta{}_{r\theta}
=\widehat\Gamma^\phi{}_{r\phi}=1/r$.
The optical screen must be constructed using the full metric,
not an equatorially reduced line element. In particular,
\begin{equation}
\label{eq:appC_mixed_derivative}
 \left.\widehat\Omega_{r\theta}\right|_{\theta=\pi/2}=0,
 \qquad
 \left.\partial_\theta\widehat\Omega_{r\theta}
 \right|_{\theta=\pi/2}=-\frac{\beta r}{\sqrt f}.
\end{equation}
The second quantity contributes to the polar curvature even though
the first vanishes.

For clarity, it is useful to perform the projection in the more
general family of Eq.~\eqref{eq:profiles_arbitrary_field}. Write
$a=a(r)>0$ and $\eta=\theta-\pi/2$. Near the common ray, the full
metric has the expansions
\begin{align}
 \widehat\Omega_{rr}
 &=-f^{-1}-\beta a^2\eta^2+O(\eta^4),\nonumber\\
 \widehat\Omega_{r\theta}
 &=-\beta ra\eta+O(\eta^3),\nonumber\\
 \widehat\Omega_{\theta\theta}
 &=-(1+\beta)r^2+\beta r^2fa^2\eta^2+O(\eta^4),
 \label{eq:appC_general_metric_expansion}
\end{align}
while $\widehat\Omega_{tt}=f$ and
$\widehat\Omega_{\phi\phi}=-r^2\sin^2\theta$.
The inverse metric has the off-diagonal expansion
$\widehat\Omega^{r\theta}=\alpha fa\eta/r+O(\eta^3)$.
The connection coefficients needed for the polar projection obey
\begin{align}
 \left.\partial_\theta\widehat\Gamma^\theta{}_{tt}
 \right|_{\pi/2}
 &=-\frac{\alpha f f'a}{2r},\nonumber\\
 \left.\partial_\theta\widehat\Gamma^\theta{}_{rr}
 \right|_{\pi/2}
 &=\frac{\alpha a f'}{2rf}
 -\frac{\alpha}{r^2}(a^2-a-ra'),
 \label{eq:appC_general_connection_derivatives}
\end{align}
where $f'=df/dr$ and $a'=da/dr$. Together with
$\widehat\Gamma^r{}_{tt}=ff'/2$,
$\widehat\Gamma^r{}_{rr}=-f'/(2f)$, and
$\widehat\Gamma^\theta{}_{r\theta}=1/r$ on the ray,
these give the exact curvature components
\begin{align}
 \widehat R^\theta{}_{tt\theta}
 &=\frac{ff'}{2r}(\alpha a-1),\nonumber\\
 \widehat R^\theta{}_{rr\theta}
 &=\frac{f'}{2rf}(1-\alpha a)
 +\frac{\alpha}{r^2}(a^2-a-ra'),\nonumber\\
 \widehat R^\phi{}_{tt\phi}
 &=-\frac{ff'}{2r},\nonumber\\
 \widehat R^\phi{}_{rr\phi}&=\frac{f'}{2rf}.
 \label{eq:appC_general_curvature_components}
\end{align}
The $tr$ cross-components appearing in the contraction with
$k^\mu k^\nu$ vanish. The polar terms proportional to $f'$ then
cancel in the screen projection:
\begin{align}
\label{eq:appC_general_curvature_projection}
 \frac{1}{f^2}\widehat R^\theta{}_{tt\theta}
 +\widehat R^\theta{}_{rr\theta}
 &=\frac{\alpha}{r^2}(a^2-a-ra'),\\
 \frac{1}{f^2}\widehat R^\phi{}_{tt\phi}
 +\widehat R^\phi{}_{rr\phi}&=0.
\end{align}
This calculation directly establishes the general profile formula
of Eq.~\eqref{eq:profiles_general_tidal}, without invoking the
neighboring-geodesic equations.

The original preferred field corresponds to $a=f^{-1/2}$. Substituting
this profile into Eq.~\eqref{eq:appC_general_curvature_components}
gives
\begin{align}
\label{eq:appC_riemann_components}
 \widehat R^\theta{}_{tt\theta}
 &=\frac{M}{r^3}(\alpha\sqrt f-f),
 &\widehat R^\theta{}_{rr\theta}
 &=\frac{M}{r^3f}
 +\frac{\alpha}{r^2}\left(\frac1f-\frac1{\sqrt f}\right),
 \nonumber\\
 \widehat R^\phi{}_{tt\phi}
 &=-\frac{Mf}{r^3},
 &\widehat R^\phi{}_{rr\phi}
 &=\frac{M}{r^3f}.
\end{align}
Consequently, the exact optical tidal matrix is
\begin{equation}
\label{eq:appC_tidal_projection}
 \mathcal T_{\theta\theta}
 =\frac1{f^2}\widehat R^\theta{}_{tt\theta}
 +\widehat R^\theta{}_{rr\theta}=\alpha F(r),
 \qquad \mathcal T_{\phi\phi}=0,
\end{equation}
with
\begin{equation}
\label{eq:appC_F}
 F(r)=\frac{M(1+3\sqrt f)}
 {r^3 f^{3/2}(1+\sqrt f)}.
\end{equation}
As required by the screen-trace identity, the Ricci contraction
and trace-free part are
\begin{equation}
\label{eq:appC_trace}
 \widehat R_{\mu\nu}k^\mu k^\nu=-\alpha F(r),
 \quad
 \mathcal T^{\rm TF}_{AB}
 =\frac{\alpha F(r)}{2}\operatorname{diag}(1,-1).
\end{equation}
For $r\gg M$, the expansion reads
\begin{equation}
\label{eq:appC_weak}
 F(r)=\frac{2M}{r^3}+\frac{11M^2}{2r^4}
 +\frac{13M^3}{r^5}
 +O\!\left(\frac{M^4}{r^6}\right).
\end{equation}
In particular, the cancellation at $\alpha=0$ concerns this
specific radial null screen projection; it does not imply that the
Schwarzschild Riemann tensor vanishes.

\subsection{Independent angular-geodesic derivation}
\label{app:tidal_geodesic_check}

As a separate check, let
$\theta=\pi/2+\eta(r)$ and $\phi=\phi_o+\zeta(r)$, where
$\eta$ and $\zeta$ are infinitesimal. At linear order, the
radial and temporal motions retain their central values,
$dr/ds=1+O(\eta^2,\zeta^2)$ and
$dt/dr=f^{-1}+O(\eta^2,\zeta^2)$.
The angular equations of the full optical geodesic system for
$a=f^{-1/2}$ are
\begin{equation}
\label{eq:appC_linear_theta}
 \eta''+\frac2r\eta'-\alpha F(r)\eta=0,
 \qquad
 \zeta''+\frac2r\zeta'=0.
\end{equation}
The orthonormal optical-screen components are
$Y_\theta=r\sqrt{1+\beta}\,\eta$ and $Y_\phi=r\zeta$.
Hence
\begin{equation}
\label{eq:appC_jacobi}
 Y_\theta''=\alpha F(r)Y_\theta,
 \qquad Y_\phi''=0,
\end{equation}
in exact agreement with Eq.~\eqref{eq:appC_tidal_projection}.
The agreement fixes the tidal sign independently of the chosen
presentation of the Riemann tensor.

\subsection{Variational check for a general radial profile}
\label{app:tidal_variational_profile}

For a general smooth positive $a(r)$, retain the polar perturbation
$\eta(r)$ at fixed azimuth. The expansion of the optical geodesic
action to second order in $\eta$ and $\eta'$ gives, up to
terms independent of the deviation and boundary terms,
\begin{align}
 L_{\perp}^{(2)}=-\frac12\bigl[&(1+\beta)r^2(\eta')^2
 +2\beta r a(r)\eta\eta'\nonumber\\
 &+\beta a(r)^2\eta^2\bigr].
 \label{eq:appC_general_quadratic}
\end{align}
Its Euler--Lagrange equation can be written before division by
the common positive factor $1+\beta$ as
\begin{equation}
 (1+\beta)(r^2\eta''+2r\eta')
 -\beta(a^2-a-ra')\eta=0.
 \label{eq:appC_general_euler_undivided}
\end{equation}
Using $\alpha=\beta/(1+\beta)$, we recover
\begin{equation}
\label{eq:appC_general_euler}
 \eta''+\frac2r\eta'
 -\frac{\alpha}{r^2}(a^2-a-ra')\eta=0.
\end{equation}
After the substitution $Y_\theta=r\sqrt{1+\beta}\,\eta$,
this becomes
\begin{equation}
 \label{eq:appC_general_jacobi}
 Y_\theta''=\frac{\alpha}{r^2}
 (a^2-a-ra')Y_\theta,
\end{equation}
reproducing the curvature-derived
Eq.~\eqref{eq:appC_general_curvature_projection}.
For $a=f^{-1/2}$ the coefficient reduces exactly to $\alpha F(r)$;
for $a=f^{-p/2}$ it yields the sign classification of
Sec.~\ref{sec:profiles}. The equality of the curvature and
variational results is independent of any field equation for $a(r)$.

\section{Volterra solution, uniform error bound, and asymptotics}
\label{app:jacobi_asymptotic}

This appendix establishes the global properties of the vertex-normalized
Jacobi solution used in Sec.~\ref{sec:beam}. We first consider the
$p=1$ background, for which $F(r)>0$ on the Schwarzschild exterior.
We distinguish the exact finite-coupling solution from its expansion
in the deformation parameter $\alpha$. We then derive the closed
asymptotic coefficient, prove existence of the large-distance limit,
and state sufficient conditions excluding conjugate points. The
extension to signed tidal profiles is considered at the end.

\subsection{Volterra representation and uniform convergence}
\label{app:volterra_convergence}

For a fixed emission radius $r_o>2M$, the polar Jacobi function obeys
\begin{equation}
\label{eq:appD_IVP}
 D_\theta''(r)=\alpha F(r)D_\theta(r),\qquad
 D_\theta(r_o)=0,\qquad D_\theta'(r_o)=1,
\end{equation}
where $F$ is the exact function in Eq.~\eqref{eq:tidal_exact_matrix}.
The azimuthal solution is $D_\phi=r-r_o$. Integrating the initial-value
problem twice gives
\begin{equation}
\label{eq:appD_volterra}
 D_\theta(r)=r-r_o+\alpha\int_{r_o}^{r}
 (r-u)F(u)D_\theta(u)\,du.
\end{equation}
For $r>r_o$, define the continuous ratio
$R(r)=D_\theta(r)/(r-r_o)$, with $R(r_o)=1$. It satisfies
\begin{align}
\label{eq:appD_ratio_volterra}
 R(r)=1+\alpha\int_{r_o}^{r}
 \frac{r-u}{r-r_o}(u-r_o)F(u)R(u)\,du.
\end{align}
The normalization agrees with the material directional-stretch ratio
of Eq.~\eqref{eq:beam_ratio}; in particular, the optical-to-material
conversion does not introduce a second deformation coefficient.

Set
\begin{equation}
\label{eq:appD_J_definition}
 \mathcal J(r_o)=\int_{r_o}^{\infty}(u-r_o)F(u)\,du.
\end{equation}
The integral is finite: $F$ is regular for $u\geq r_o$ and
$F(u)=2M/u^3+O(u^{-4})$ as $u\to\infty$. Moreover,
\begin{equation}
\label{eq:appD_kernel_majorant}
 0\leq\frac{r-u}{r-r_o}\leq1
 \qquad (r_o\leq u\leq r),
\end{equation}
so the Volterra kernel is dominated by the integrable,
nonnegative weight $w(u)=(u-r_o)F(u)$.

Write the Neumann series as
$R(r)=\sum_{n=0}^{\infty}\alpha^n a_n(r)$, with $a_0=1$.
For the positive $p=1$ kernel, its coefficients satisfy
$0\leq a_n(r)$, and the ordered-simplex estimate gives
\begin{equation}
\label{eq:appD_neumann_bound}
 0\leq a_n(r)\leq\frac{1}{n!}
 \left[\int_{r_o}^{r}w(u)\,du\right]^n
 \leq\frac{\mathcal J(r_o)^n}{n!}.
\end{equation}
It follows that the series converges absolutely and uniformly for
$r\in[r_o,\infty)$ for every finite real $\alpha$, and
\begin{equation}
\label{eq:appD_ratio_bound}
 |R(r)|\leq\exp\bigl(|\alpha|\mathcal J(r_o)\bigr).
\end{equation}
Since $a_1(r)=\mathcal I(r;r_o)$, where
\begin{equation}
\label{eq:appD_I}
 \mathcal I(r;r_o)=\frac{1}{r-r_o}
 \int_{r_o}^{r}(r-u)(u-r_o)F(u)\,du,
\end{equation}
the remainder after first order obeys
\begin{align}
\label{eq:appD_error}
 \left|R(r)-1-\alpha\mathcal I(r;r_o)\right|
 \leq e^{|\alpha|\mathcal J(r_o)}-1
 -|\alpha|\mathcal J(r_o).
\end{align}
The estimate is uniform in the reception radius and holds at the
vertex by continuity. The series thus converges for finite $\alpha$,
whereas a quantitatively accurate first-order truncation requires
$|\alpha|\mathcal J(r_o)\ll1$. These statements are not equivalent.

\subsection{Evaluation of the asymptotic response coefficient}
\label{app:volterra_coefficient}

The first-order response has a finite limit because
$0\leq\mathcal I(r;r_o)\leq\mathcal J(r_o)$ and dominated
convergence gives $\mathcal I(r;r_o)\to\mathcal J(r_o)$ as
$r\to\infty$. To evaluate this coefficient, introduce
\begin{equation}
\label{eq:appD_z_change}
 z=\sqrt{1-\frac{2M}{u}},\qquad
 z_o=\sqrt{1-\frac{2M}{r_o}},\qquad
 u=\frac{2M}{1-z^2}.
\end{equation}
Substitution of the exact $F(u)$ into
Eq.~\eqref{eq:appD_J_definition} yields
\begin{equation}
\label{eq:appD_J_quadrature}
 \mathcal J(r_o)
 =\int_{z_o}^{1}\frac{(z^2-z_o^2)(1+3z)}
 {(1-z_o^2)z^2(1+z)}\,dz.
\end{equation}
The integrand can be decomposed into elementary rational terms,
\begin{align}
\label{eq:appD_partial_fractions}
 \frac{(z^2-z_o^2)(1+3z)}
 {(1-z_o^2)z^2(1+z)}
 ={}&\frac{3}{1-z_o^2}-\frac{2}{1+z}\nonumber\\
 &-\frac{2z_o^2}{(1-z_o^2)z}
 -\frac{z_o^2}{(1-z_o^2)z^2}.
\end{align}
Integration gives the closed result
\begin{align}
\label{eq:appD_J_closed}
 \mathcal J(r_o)={}&\frac{3-z_o}{1+z_o}
 +2\ln\!\left(\frac{1+z_o}{2}\right)\nonumber\\
 &+\frac{2z_o^2}{1-z_o^2}\ln z_o,
 \qquad 0<z_o<1.
\end{align}
Its monotonicity and endpoint values follow directly from the
positive integral representation:
\begin{align}
\label{eq:appD_J_derivative}
 \frac{d\mathcal J}{dr_o}
 &=-\int_{r_o}^{\infty}F(u)\,du<0,\\\label{eq:appD_Jlimits}
 \lim_{r_o\to\infty}\mathcal J(r_o)&=0,\\
 \lim_{r_o\to2M^+}\mathcal J(r_o)&=3-2\ln2.
\end{align}
In particular,
$0<\mathcal J(r_o)<3-2\ln2$ for every $r_o>2M$.
Expansion at fixed $M$ for $r_o\gg M$ produces
\begin{align}
\label{eq:appD_J_weak}
 \mathcal J(r_o)={}&\frac{M}{r_o}
 +\frac{11}{12}\left(\frac{M}{r_o}\right)^2
 +\frac{13}{12}\left(\frac{M}{r_o}\right)^3\nonumber\\
 &+\frac{233}{160}\left(\frac{M}{r_o}\right)^4
 +O\!\left(\frac{M^5}{r_o^5}\right).
\end{align}
The finite limit as $r_o\to2M^+$ concerns the exterior integral;
it is not a claim that static observers or the prescribed preferred
field admit a regular continuation across the horizon.

\subsection{Exact large-distance limit and logarithmic correction}
\label{app:volterra_large_distance}

The uniform bound implies
$|D_\theta(r)|\leq(r-r_o)e^{|\alpha|\mathcal J(r_o)}$.
Since $F(r)D_\theta(r)=O(r^{-2})$, integrating
Eq.~\eqref{eq:appD_IVP} once establishes the finite limit
\begin{equation}
\label{eq:appD_Ainfty}
 A_\infty\equiv\lim_{r\to\infty}D_\theta'(r)
 =1+\alpha\int_{r_o}^{\infty}F(u)D_\theta(u)\,du.
\end{equation}
The integral equation and the Ces\`aro limit then give
\begin{equation}
\label{eq:appD_ratio_infty}
 \lim_{r\to\infty}\frac{D_\theta(r)}{r-r_o}
 =A_\infty=1+\alpha\mathcal J(r_o)+O(\alpha^2).
\end{equation}
More precisely, Eq.~\eqref{eq:appD_error} yields
\begin{align}
\label{eq:appD_asymptotic_error}
 |A_\infty-1-\alpha\mathcal J(r_o)|
 \leq e^{|\alpha|\mathcal J(r_o)}-1
 -|\alpha|\mathcal J(r_o).
\end{align}
The formula for $A_\infty$ is exact; the expression involving
$\mathcal J$ is its first-order expansion, not an exact finite-coupling
solution.

The far-field expansion of $F$ further determines the first
subleading term of the Jacobi solution. From
$F(r)=2M/r^3+O(r^{-4})$ and $D_\theta(r)=A_\infty r+O(\ln r)$,
one obtains
$D_\theta''(r)=2\alpha M A_\infty/r^2
+O(\ln r/r^3)$. Integration from infinity gives
\begin{align}
\label{eq:appD_large_r}
 D_\theta(r)={}&A_\infty r
 -2\alpha M A_\infty\ln(r/M)+B_\infty
 \nonumber\\
 &+O\!\left(\frac{\ln(r/M)}{r}\right),
\end{align}
with a finite constant $B_\infty$ depending on the vertex data.
Thus the logarithmic correction is subleading in the directional
ratio $D_\theta/(r-r_o)$.

\subsection{Conjugate-point bounds and signed profiles}
\label{app:volterra_caustics}

For $\alpha>0$, positivity of $F$ and the initial conditions imply
$D_\theta(r)>0$ and $D_\theta''(r)>0$ for $r>r_o$. Hence
$D_\theta(r)>r-r_o$, and no conjugate point occurs beyond the vertex.
For $\alpha<0$, suppose a first zero occurs at $r_c>r_o$.
Before that zero, $D_\theta$ is positive and concave;
therefore $0<D_\theta(u)\leq u-r_o$ for $r_o<u<r_c$.
Evaluating the Volterra equation at the first zero gives
\begin{align}
\label{eq:appD_no_caustic}
 0=D_\theta(r_c)
 &\geq(r_c-r_o)\bigl[1-|\alpha|\mathcal J(r_o)\bigr].
\end{align}
Consequently,
\begin{equation}
\label{eq:appD_no_caustic_sufficient}
 |\alpha|\mathcal J(r_o)<1
\end{equation}
is a sufficient, not necessary, condition excluding conjugate points.
Using the uniform bound on $\mathcal J$ and
$\alpha=\kappa_A/(1+\kappa_A/2)$, a source-radius-independent
sufficient condition for negative coupling is
\begin{equation}
\label{eq:appD_kappa_bound}
 -\frac{2}{7-4\ln2}<\kappa_A<0.
\end{equation}
Its mathematical validity in the optical model does not resolve
causal-admissibility questions relative to conventional matter.

For a general signed tidal coefficient $F_p(r)$ with
\begin{equation}
\label{eq:appD_K_signed}
 \mathcal K_p(r_o)=\int_{r_o}^{\infty}
 (u-r_o)|F_p(u)|\,du<\infty,
\end{equation}
the same ordered-simplex argument, now applied to absolute values,
gives the uniform remainder bound
\begin{align}
\label{eq:appD_signed_error}
 |\mathcal R_p(r;r_o)-1-\alpha\mathcal I_p(r;r_o)|
 \leq e^{|\alpha|\mathcal K_p(r_o)}
 -1-|\alpha|\mathcal K_p(r_o).
\end{align}
The signed first-order integral $\mathcal J_p$ does not replace
$\mathcal K_p$ in an absolute error estimate when $F_p$ changes sign.

The $p=-1$ example of Sec.~\ref{sec:profiles} has $F_{-1}<0$.
Its coefficient is given exactly by
Eq.~\eqref{eq:profiles_J_minus_one}, with
\begin{equation}
\label{eq:appD_pminus_bound}
 0<\mathcal K_{-1}(r_o)=-\mathcal J_{-1}(r_o)
 <2\ln2-\frac56<1.
\end{equation}
For $0<\kappa_A<2$, one has $0<\alpha<1$; the polar equation
then has a nonpositive tidal potential. If a first positive-radius
zero existed, the same concavity argument would require
$|\alpha|\mathcal K_{-1}(r_o)\geq1$, contradicting
Eq.~\eqref{eq:appD_pminus_bound}. Thus $D_\theta$ remains
strictly positive for $p=-1$ and positive coupling despite its
compressive tidal contribution.

These conclusions concern the vertex-normalized linear Jacobi
problem along the prescribed exterior ray. They do not provide a
nonlinear finite-aperture optical map, nor do they establish a
causal or dynamical completion of the chosen Lorentz-violating
background.

\bibliographystyle{apsrev4-2}
\bibliography{refs_cqg_consolidadas}

\end{document}